\documentclass[pra, 
aps,
amsmath,
amssymb,
twocolumn,
nofootinbib,
superscriptaddress,
]{revtex4-2}
\usepackage{graphicx,dcolumn,bm}
\usepackage{amsfonts,amsmath,amssymb,amsthm,amscd}
\usepackage[english]{babel}
\usepackage{physics}
\usepackage{needspace}
\usepackage{xcolor}
\usepackage{bm}
\definecolor{linkblue}{HTML}{2e3092}
\usepackage[colorlinks=true, allcolors = linkblue]{hyperref}
\renewcommand{\vec}[1]{\bm{#1}}
\AtBeginDocument{\renewcommand{\natexlab}[1]{#1}}

\begin{document}
\title{Onset of electron-seeded cascades in generic electromagnetic fields}
\author{A.~A. Mironov}\email{mironov.hep@gmail.com}
\affiliation{Prokhorov General Physics Institute of the Russian Academy of Sciences, Moscow, 119991, Russia}
\affiliation{National Research Nuclear University MEPhI, Moscow, 115409,  Russia}
\author{E.~G. Gelfer}\email{egelfer@gmail.com}
\affiliation{ELI Beamlines, Institute of Physics of the ASCR, v.v.i., Dolni Brezany, Czech Republic}
\author{A.~M.~Fedotov}\email{am\_fedotov@mail.ru}
\affiliation{National Research Nuclear University MEPhI, Moscow, 115409,  Russia}
\affiliation{Laboratory for Quantum Theory of Intense Fields, National Research Tomsk State University, Tomsk, 634050, Russia}

\begin{abstract}
QED cascades in a strong electromagnetic field of optical range and arbitrary configuration are considered. A general expression for short-time dependence of the key electron quantum dynamical parameter is derived, allowing to generalize the effective threshold condition of QED cascade onset. The generalized theory is applied to selfsustained cascades in a single focused laser pulse. According to numerical simulations, if a GeV electron bunch is used as a seed, an ordinary cascade can be converted into the selfsustained one.  As an application, it would be also possible to produce this way bright collimated photon beams with up to GeV photon energies.
\end{abstract}

\maketitle

\section{Introduction}

Recent advances in developing high-power lasers (see e.g.  the recent review \cite{danson2019petawatt}) offer impressive prospects for bringing the experimental studies of laser-matter interaction to a new level \cite{di2012extremely}.  It was recently reported that laser intensity of $10^{23}$ W/cm$^2$ was exceeded at the CoReLS facility \cite{yoon2021one}, and the new multi-petawatt-class facilities under construction (Vulcan2020 \cite{hernandez2010vulcan}, PEARL 10 \cite{korzhimanov2011horizons,shaykin2014prospects}, Apollon 10PW  \cite{papadopoulos2017high}, ELI Beamlines \cite{a_ELI,weber2017p3}, ELI-NP \cite{ELI-NP,tanaka2020current}, SULF \cite{gan2017200}, SEL \cite{cartlidge2018light}, etc.) are promising in probing intensities $\simeq 10^{23-24}$ W/cm$^2$ in the nearest future. Furthermore, far-reaching exawatt-class facilities, such as ELI Fourth Pillar \cite{mourou2011eli} and/or XCELS \cite{shaykin2014prospects,XCELS}, aiming at intensity $\gtrsim 10^{26}$ W/cm$^2$, are also being planned. One of the main possible applications would be study of a variety of QED phenomena that have been never observed previously \cite{di2012extremely,narozhny2014creation,narozhny2015extreme,blackburn2020radiation,hu2020seed,zhang2020relativistic}. 

Selfsustained [or A(valanche)-type]  QED cascades, the primary topic of the paper, are the examples of such phenomena. They are long chains of successive non-linear Compton scatterings (hard photon emission) and Breit-Wheeler (pair photoproduction) processes. After repartitioning through these processes, the energy of secondary particles is each time reimbursed due to their ongoing acceleration by laser field. This results in a rapid development of an avalanche of up to macroscopic multiplicity, with the energy source provided and hence the parameters determined by the field, rather than by seed particles (kind of losing memory of initial conditions). 

Onset of selfsustained cascades in interactions of extremely high intensity laser pulses with matter was predicted theoretically \cite{bell2008possibility, kirk2009pair, fedotov2010limitations}, initially by implying a particular setup with a seed electron placed into a common focus of counter-propagating laser pulses. Variations of that setup combining two or more laser pulses were studied by numerical simulations in a number of subsequent papers, see e.g. Refs.~\cite{duclous2010monte, nerush2011analytical, nerush2011laser, king2013photon, bashmakov2014effect,gelfer2015optimized, gelfer2016generation, grismayer2017seeded, seipt2020polarized}. As a rule, the simulations were initialized starting with a rest seed electron at the focus, thus leaving aside a non-trivial problem of injecting charged particles into a strong field region \cite{elkina2014improving,fedotov2014radiation}. Further simulations, however, demonstrated generation of cascades in more realistic settings, e.g. by irradiating dilute gas or electron cloud \cite{jirka2016electron,tamburini2017laser,gonoskov2017ultrabright,artemenko2017ionization}, near-critical-density plasma \cite{zhu2016dense,zhu2018generation}, or dense solid \cite{ridgers2012dense, kirk2013pair,jirka2017qed,slade2019identifying, samsonov2019laser, samsonov2020hydrodynamical} targets. Note that in the latter case it was enough to irradiate a target from one side, since due to reflection a counterpropagating wave was formed by itself.

Another conceivable option is using a multi-GeV electron bunch as a target \cite{sokolov2010pair,bulanov2013electromagnetic,mironov2014collapse, mironov2016generation,blackburn2017scaling,vranic2018multi,magnusson2019multiple,wan2020ultrarelativistic}. The idea was first implemented in the SLAC E-144 experiment \cite{bula1996observation,burke1997positron,bamber1999studies}, where 46.6 GeV electron bunches collided with  laser pulses focused to intensity $\sim 10^{18}$ W/cm$^2$. However, due to the low intensity of the laser pulse production of only few pairs (rather than of a cascade chain) could be detected. The two upcoming experiments E-320 at SLAC \cite{meuren2020seminal} and LUXE at DESY \cite{abramowicz2019letter,abramowicz2021conceptual} will advance the SLAC E-144 setup by bringing the laser field intensity up to the level of $\sim 10^{20}$ W/cm$^2$.

We call a cascade S(hower)-type if $e^+e^-$ pairs and hard photons are produced at the expense of kinetic energy of the initial and secondary particles. Such a cascade eventually saturates as soon as the energy of secondary particles drops to a threshold value. Note that after the saturation of the S-type cascade an A-type cascade can set in (the so-called `cascade collapse and revival' effect). According to simulations \cite{mironov2014collapse, mironov2016generation}, cascade `revival' requires laser intensity $10^{23}-10^{24}$ W/cm$^2$ and certain tweaking of the electron bunch parameters.

As reported by now, most of simulations of cascade production assumed variations of a particular setup with two or more incoming laser pulses. This was partially because of the intuitive expectations supported by numerical simulations, that the effective threshold intensity for cascade onset in such a scheme is lower, thus setting up the problem closer to the existing experimental capabilities. However, a more technical reason was the existence of a simple qualitative theory \cite{fedotov2010limitations,nerush2011analytical,kostyukov2018growth}. It is essentially based on replacing the actual (generally rather complicated and in reality unknown) field configuration in a laser focus by a `uniformly rotating electric field'. Existence of such a pivotal theory greatly simplified codes benchmarking and interpretation of the simulation results. Though some attempts have been already made \cite{bashmakov2014effect,gelfer2015optimized,grismayer2017seeded} towards generalization, they still essentially relied on the assumption that the magnetic field vanishes at the electric field maximum, thus ruling out the important case of a single focused laser pulse in vacuum.

Our goal below is to derive a qualitative criterion of an A-type cascade onset for initially nonresting seed particles in an arbitrary field configuration. Our present approach is completely based on a formal short-time expansion of electron motion in an arbitrary field of ultrarelativistic intensity, thus an appeal to oversimplified field models is no longer required. In particular, we show explicitly that any slowly varying electromagnetic (EM) field of electric type and sufficient magnitude is capable for A-type cascade production. 

We apply our theory to the case of cascade production by a single focused laser pulse. This problem is related to the possible fundamental limitations on the intensity attainable with high power lasers \cite{fedotov2010limitations}. Recall that all the existent proposals for increasing laser intensity rely in that or other way on employing tight focusing. But it is known \cite{narozhny2004e+} that a focused field is always of electric ($E>H$) type in certain spatial regions, hence at sufficiently high intensity will unavoidably create pairs from vacuum. The spontaneously produced pairs should in turn seed massive A-type QED cascades capable for depletion of the laser field, as it was explicitly demonstrated for a setup with two counterpropagating laser pulses in Refs.~\cite{nerush2011laser,grismayer2016laser}, thus imposing limitations on the attainable laser intensity. Obviously, an upper bound on the attainable intensity corresponds to the worst case of the maximal possible threshold for cascade production, that is of a single focused pulse. 

Here we study analytically and numerically the dependence of the effective threshold intensity upon the degree of focusing. It is shown that even though (as anticipated) for moderate focusing the threshold intensity for A-type cascade production by a single focused pulse is typically several orders higher than for two, not to mention more, colliding pulses, it nevertheless still remains lower than the threshold for spontaneous pair production, thus promoting the latter as an upper bound for the attainable laser intensity in vacuum. 

We also study cascades arising in a head-on collision of a GeV electron beam with a focused laser pulse. Quantum radiation \cite{blackburn2014quantum} in such a scenario, as well as S-type cascades for either near-presently available or higher intensity without account for focusing (plane wave case) have been discussed previously \cite{bulanov2013electromagnetic,sokolov2010pair}. However, according to our simulations, by appropriate tuning of the parameters, the effect of `collapse and revival' (i.e. S-type to A-type cascade conversion) can take place as well. 

A byproduct of setting up cascades with a single laser pulse is production of collimated gamma-rays with specific properties. Cascades produced by irradiating solid targets with high intensity laser pulses are known to effectively convert laser  energy into hard gamma quanta \cite{ridgers2013dense, nerush2014gamma, ji2014energy, brady2014synchrotron, bashinov2014new,gu2018brilliant,a_zhu2018bright}. An A-type cascade arising at interaction of a GeV electron beam with a focused laser pulse of intensity $I\sim 10^{26}$ W/cm$^2$ should also serve as a bright source of collimated (emitted towards the laser pulse propagation direction) GeV photons. But, unlike the previously considered schemes, here collimation is achieved automatically and the parameters of the resulting gamma pulses are controlled exclusively by the driving laser pulse.

The paper is organized as follows. In Sec.~\ref{sec:basic} we introduce the notations and give an overview of the general approximations in use. In Sec.~\ref{sec:chi(t)} we consider classical motion of a seed electron in an arbitrary EM field with a goal of deriving a short-time dependence for its energy and dynamical quantum parameter. Based on these results, we derive and discuss general criteria for a selfsustained cascade onset in Sec.~\ref{sec:criteria}. In Sec.~\ref{sec:sp_estimates} we study an electron-seeded cascade in a single focused laser pulse by applying both the developed theory and numerical simulations. In the follow-up Sec.~\ref{sec:collapse_and_revival} we present the simulations demonstrating that such a cascade indeed develops in a  head-on collision of a GeV electron beam with a single laser pulse. Furthermore, we discuss the angular and energy distributions of the gamma-quanta produced throughout the collision. Summary of our results and the concluding remarks are collected in Sec.~\ref{sec:summary}. Appendix~\ref{sec:app_rot} illustrates general arguments of Sec.~\ref{sec:chi(t)} with an example of electron dynamics in a uniformly rotating electric field. Finally, Appendix~\ref{sec:app_sp} contains technical details of our derivations for a single focused laser pulse.

\section{General theory of A-type cascades}
\label{sec:general}
\subsection{Basic assumptions}
\label{sec:basic}

A QED cascade is a chain of the successive first-order QED processes, the nonlinear Compton scattering (hard photon emission) and the multiphoton Breit-Wheeler process (pair photoproduction). Laser field is characterized by an invariant dimensionless parameter ~\footnote{We use units $\hbar=c=1$.} $a_0=e\sqrt{-A^\mu A_\mu}/m$, where $A_\mu$ is the 4-potential and $e$ and $m$ denote the magnitude of the electron charge and its mass, respectively. We assume the field is of (near-)optical frequency and ultrarelativistic intensity ($a_0\gg 1$). Under such conditions the field can be considered as slowly varying over the characteristic formation scales of QED processes. This allows to apply the locally constant field approximation (LCFA) \cite{ritus1979fian111,fedotov2010limitations,elkina2011qed,mironov2016generation} by using the probability rates of the QED processes in a constant crossed field \cite{nikishov1964quantum,nikishov1967pair,baier1967quantum}. They are determined by the dimensionless dynamical quantum parameter \cite{ritus1979fian111} 
\begin{equation}
\label{eq:chi}
\chi_{e,\gamma} = \frac{e}{m^3}\sqrt{\left(p_0\vec{E}+\vec{p}\times\vec{H}\right)^2-\left(\vec{p}\cdot\vec{E}\right)^2}
\end{equation}
of the involved particle, where $\vec{p}$ is its momentum and $\vec{E}$, $\vec{H}$ are the electric and magnetic fields at the particle position. Strictly speaking, it is known that LCFA is valid if $a_0\gg\max(1,\chi^{1/3})$ and, additionally, for emission of not too soft photons (namely, only those with $\chi_\gamma\gg\chi_e^2/a_0^3$) \cite{ritus1979fian111,meuren2016semiclassical,blackburn2018benchmarking,ilderton2019extended,raicher2019semiclassical, di2019improved,heinzl2020locally,king2020uniform,blackburn2020radiation}. We assume that these conditions are fulfilled and are interested only in emission of photons with $\chi_{\gamma}\gtrsim 1$.

The total probability rates $W_\text{rad}$ and $W_\text{cr}$ for hard photon emission and pair photoproduction, respectively, admit an especially simple form in the asymptotic cases \cite{ritus1979fian111,elkina2011qed}:
\begin{subequations}
\begin{equation}
\label{eq:W_rad}
W_\text{rad}\approx \frac{\alpha m^2}{\varepsilon_e}\times\left\lbrace
\begin{array}{ll}
1.44\,\chi_e, & \chi_e\ll 1, \\
1.46\,\chi_e^{2/3}, & \chi_e\gg 1,
\end{array}\right.
\end{equation}
\begin{equation}
\label{eq:W_cr}
W_\text{cr}\approx \frac{\alpha m^2}{\varepsilon_\gamma} \times \left\lbrace
\begin{array}{ll}
0.23\,\chi_\gamma e^{-8/3\chi_\gamma}, & \chi_\gamma\ll 1, \\
0.48\,\chi_\gamma^{2/3}, & \chi_\gamma\gg 1,
\end{array}\right.
\end{equation}
\end{subequations}
where $\varepsilon_{e,\gamma}$, $\chi_{e,\gamma}$ are, respectively, the energy and the dynamical quantum parameter (\ref{eq:chi}) of an incoming electron (positron) or photon, and $\alpha$ is the fine structure constant. In particular, pair photoproduction by soft photons is exponentially suppressed, this is the reason not to focus on their emission carefully.

At the same time, we assume that $E,H\ll E_S$, where $E_S=m^2/e=1.32\times 10^{16}$ V/cm is the QED critical field \cite{sauter1931verhalten,schwinger1951gauge}, and that the majority of particles in a cascade are most of the time ultrarelativistic. Then a semiclassical approach is valid, which treats the charged particles and hard photons pointlike and propagating along the classical trajectories in between the events of photon emission or pair photoproduction. The QED processes mentioned above are included by using a Monte Carlo event generator implementing the known probability rates as described in detail in Refs.~\cite{elkina2011qed,mironov2016generation}. We emphasize that the particle trajectories are determined by the Lorentz equations, with no need to correct them by adding a classical radiation friction force, which is already  included as a quantum radiation recoil at the level of the Monte Carlo event generator, see the Appendix in Ref.~\cite{elkina2011qed}.

\subsection{Short-time dependence $\chi(t)$ in generic field}
\label{sec:chi(t)}

A key distinctive feature of an A-type cascade is the ongoing reimbursement of the energy and the dynamical quantum parameter (\ref{eq:chi}) of the participating charged particles due to their acceleration in the field. Let us study it in general setting. Suppose an electron is placed into a strong field region. Within a semiclassical approach its trajectory is governed by the equations of motion
\begin{subequations}
\begin{equation}\label{eq:eq_motion}
\frac{dp^\mu(\tau)}{d\tau}=\frac{e}{m}F^\mu_{\;\;\;\nu}(x(\tau))p^\nu(\tau),
\end{equation}
\begin{equation}\label{eq:eq_motion1}
\quad \frac{dx^\mu(\tau)}{d\tau}=\frac{p^\mu(\tau)}{m},
\end{equation}
\end{subequations}
supplemented (without a loss of generality) with the initial conditions at $\tau=0$:
\begin{equation}
\label{eq:in_cond}
x^\mu(0)=0,\quad p^\mu(0)=p_0^\mu\equiv\{\varepsilon_0,\vec{p}_0\},
\end{equation}
where $\tau$ is the proper time, $x^\mu=\{t,\vec{x}\}$ and $p^\mu=\{\varepsilon,\vec{p}\}$ are the 4-vectors of particle position and momentum, and $F_{\mu\nu}$ is the EM field tensor. Having a solution to these equations, it is enough to substitute it directly into Eq.~(\ref{eq:chi}) to obtain $\chi(t)$. But obviously, this cannot be done explicitly for an arbitrary field configuration. However, two aspects simplify the task: (i) we assume the field varies slowly; and (ii) we assume the field is of ultrarelativistic intensity ($a_0\gg 1$), thus it shortly makes charged particles ultrarelativistic.

In view of (i) above, let us write $F_{\mu\nu}\equiv F_{\mu\nu}(\omega x)$ and expand the particle trajectory in powers of the field carrier frequency $\omega$:
\begin{subequations}
\begin{align}
\label{eq:basic_exp_p}
p(\tau)=p^{(1)}+p^{(2)}+\ldots, \quad p^{(i)}=\mathcal{O}\left(\omega^{i-1}\right),\\
\label{eq:basic_exp_x}
x(\tau)=x^{(1)}+x^{(2)}+\ldots,\quad x^{(i)}=\mathcal{O}\left(\omega^{i-1}\right),
\end{align}
where $p^{(i)}$ and $x^{(i)}$ are the functions of $\tau$. In what follows, we will see that the actual (small) expansion parameter is $\omega t$, where $t=x^0$ is the laboratory frame time. In order to fulfil the initial conditions [see Eq.~\eqref{eq:in_cond}], we impose them termwise as follows,
\[
p^{(1)}(0)=p_0,\quad p^{(i>1)}(0)=0,\quad x^{(i)}(0)=0.
\]
Accordingly, the field expands as
\begin{equation}\label{eq:exp_F}
\left.F_{\mu\nu}(\omega x)\right|_{x=0}=F^{(0)}_{\mu\nu}+ F^{(1)}_{\mu\nu}+ F^{(2)}_{\mu\nu}+\ldots,
\end{equation}
\end{subequations}
where
\begin{equation}
\label{eq:basic_exp_F}
\begin{array}{l}
F^{(0)}_{\mu\nu} = F_{\mu\nu}\left(0\right)=\mathcal{O}(1),\\
F^{(1)}_{\mu\nu} = F_{\mu\nu,\sigma}\left(0\right)x^{(1)\,\sigma}=\mathcal{O}(\omega),\\
\begin{split}
F^{(2)}_{\mu\nu} = &\tfrac12 F_{\mu\nu,\sigma\rho}\left(0\right)x^{(1)\,\sigma}x^{(1)\,\rho} \\&
+F_{\mu\nu,\sigma}\left(0\right)x^{(2)\,\sigma}=\mathcal{O}\left(\omega^2\right),\ldots
\end{split}
\end{array}
\end{equation}
By squaring expansion  (\ref{eq:basic_exp_p}) and collecting the terms of the same order in the on-shell condition $p^2=m^2$, we arrive at the constraints
\begin{subequations}
\begin{eqnarray}
\label{eq:id1}
\left(p^{(1)}\right)^2=m^2,\\
\label{eq:id2}
p^{(1)}p^{(2)}=0,\\
\label{eq:id3}
\left(p^{(2)}\right)^2+2p^{(1)}p^{(3)}=0,\\
\nonumber\ldots
\end{eqnarray}
\end{subequations}

Next, by substituting expansions (\ref{eq:basic_exp_x})-(\ref{eq:exp_F}), Eqs.~ (\ref{eq:eq_motion}) can be solved successively up to a desired order. Let us find a solution to the least order contributing non-trivially to Eq.~(\ref{eq:chi}). Matching the terms of the order $\mathcal{O}\left(\omega\right)$, we obtain
\begin{equation}
\label{eq:1_order}
\frac{dp^{(1)}}{d\tau}-\frac{e}{m}F^{(0)}p^{(1)}=0.
\end{equation}
This is equation of motion of a charged particle in a \textit{constant}  field. For brevity, here and below we present equations in a matrix form using the abbreviations $p\equiv p^\mu$ and $F\equiv F^\mu_{\;\;\;\nu}$.

The solution for Eq.~(\ref{eq:1_order}) can be written in a matrix form (see e.g. \cite{taub1948orbits}):
\begin{equation}
\label{eq:1_order_sol}
 p^{(1)}(\tau)=e^{eF^{(0)}\tau/m}p_0=
\sum_{q=1}^4 e^{e\lambda_q\tau/m}C_q f_q,
\end{equation}
where $\lambda_q$ and $f_q$ are the eigenvalues and eigenvectors of the matrix $F^{(0)}$ defined by the equation $F^{(0)}f_q=\lambda_q f_q$. Its four solutions
\begin{equation}
\label{eq:eigenvals}
\lambda_q=\{\epsilon,-\epsilon,i\eta,-i\eta\}
\end{equation}
are expressed through the field invariants $\epsilon,\,\eta=\sqrt{\sqrt{\mathcal{F}^2+\mathcal{G}^2}\pm\mathcal{F}}$, where $\mathcal{F}=\left(\vec{E}^2-\vec{B}^2\right)/2$ and $\mathcal{G}=\vec{E}\cdot\vec{B}$ \cite{landau2013classical}. The constants $C_q$ are to be determined to provide $\sum_q C_q f_q=p_0$ according to the initial condition. For a field of electric type ($\mathcal{F}>0$) we have also $\epsilon>0$, hence on a time scale $\tau\gg m/e\epsilon$ (when the initially resting particle becomes ultrarelativistic, or, more generally, the energy and momentum acquired from the field exceed the initial ones) the major contribution in (\ref{eq:1_order_sol}) comes from the term with $q=1$ (as the rest terms are exponentially suppressed as compared to that one), so that Eq.~(\ref{eq:1_order_sol}) simplifies to
\begin{subequations}
\begin{equation}
\label{eq:ultrarel}
p^{(1)}(\tau)\approx e^{e\epsilon\tau/m}C_1 f_1,\quad \tau\gg \frac{m}{e\epsilon}.
\end{equation}
Then, by integrating Eq.~(\ref{eq:eq_motion1}), we further obtain 
\begin{equation}
\label{eq:ultrarel1}
x^{(1)}(\tau) \approx \frac1{e\epsilon}e^{e\epsilon\tau/m}C_1f_1,\quad \tau\gg \frac{m}{e\epsilon}.
\end{equation}
\end{subequations}
Assuming\footnote{This is possible for any field of electric type, since due to antisymmetry of $F^{(0)}$ we have ${\rm Re}\,(\lambda)\cdot f_\mu^*f^\mu=0$, hence for any real nonzero $\lambda=\epsilon>0$ we have $|f^0|=\left\vert\vec{f}\right\vert\ne 0$ (or else $f^\mu\equiv 0$ was not an eigenvector). Note that with such a normalization all the constants $C_q$ acquire the dimension of mass. Note also that with the adopted normalization $f^\mu$ (unlike $x^{(1)\mu}=f_1^\mu t$, see below) does not transform as a 4-vector under Lorentz transformations.} normalization $f_1^0=1$, by expressing Eqs.~(\ref{eq:ultrarel})-(\ref{eq:ultrarel1}) in terms of the laboratory time $t=x^0\approx(x^0)^{(1)}$, we have $\vec{x}^{(1)}(t)\approx \vec{f}_1 t$ and $p^{(1)}(t)\approx e\epsilon t f_1$. In particular, the electron energy on this time scale grows as
\begin{equation}
\label{eq:ultrarel_p}
\varepsilon(t)\equiv p^{0}(t)\approx e\epsilon t.
\end{equation}
At this stage the tentative conditions of validity of the approximation under development can be summarized by a double constraint
\begin{equation}
\label{eq:t_valid}
\frac{C_1}{e\epsilon}\ll t \ll \frac{\pi}{\omega}.
\end{equation}
It is consistent in strong fields $a_0\gg 1$ and defines explicitly the timescale of interest in this Section. At this timescale the electrons are ultrarelativistic, still the field varies slowly.

Balancing the terms of the next few orders of the expansions (\ref{eq:basic_exp_x})-(\ref{eq:exp_F}) in Eq.~(\ref{eq:eq_motion}), we obtain for the $\mathcal{O}\left(\omega^2\right)$-terms:
\begin{subequations}
\begin{equation}
\label{eq:2_order}
\frac{dp^{(2)}}{d\tau}-\frac{e}{m}F^{(0)}p^{(2)}=\frac{e}{m}F^{(1)}p^{(1)},
\end{equation}
and for the $\mathcal{O}\left(\omega^3\right)$-terms:
\begin{equation}
\label{eq:3_order}
\frac{dp^{(3)}}{d\tau}-\frac{e}{m}F^{(0)}p^{(3)}=\frac{e}{m}\left(F^{(1)}p^{(2)}+F^{(2)}p^{(1)}\right).
\end{equation}
\end{subequations}
It is clear from (\ref{eq:basic_exp_F}), (\ref{eq:ultrarel}) and (\ref{eq:ultrarel1}) that the RHS in Eq.~(\ref{eq:2_order}) is proportional to $x_\mu^{(1)}p_\nu^{(1)}\propto e^{2e\epsilon \tau /m}$. Hence $p^{(2)}$ should be asymptotically (for $\tau\gg m/e\epsilon$) also $\propto e^{2e\epsilon \tau /m}\propto t^2$, in particular implying that
\begin{equation}\label{eq:dp2}
\frac{dp^{(2)}}{d\tau}\approx \frac{2e\epsilon}{m}p^{(2)}.
\end{equation}
Thus the solution to Eq.~(\ref{eq:2_order}) can be written as
\begin{equation}
\label{eq:p2}
\quad p^{(2)}\approx\frac1{2\epsilon-F^{(0)}}F^{(1)}p^{(1)},
\end{equation}
where division by a matrix means a product with its inverse.  Note that with Eq.~(\ref{eq:p2}) we can verify (\ref{eq:id2}) directly by noting that $p^{(1)}$ is proportional to the eigenvector $f_1$ of $F^{(0)}$, and recognizing that $p^{(1)}F^{(1)}p^{(1)}=0$ since $F^{(1)}$ is antisymmetric. Strictly speaking, according to Eqs.~\eqref{eq:eq_motion1} and \eqref{eq:basic_exp_x}, $t$ is also a subject to second and higher order corrections. However, we are interested in the leading order nontrivial contribution to $\chi(t)$, therefore neglect all such corrections. 
 
The third order correction $p^{(3)}$ is derived from (\ref{eq:3_order}) in the same manner by noticing that in ultrarelativistic approximation $p^{(3)}\propto e^{3e\epsilon\tau/m}\propto t^3$. But, as we demonstrate below, dealing with its explicit expression can be avoided by applying Eq.~(\ref{eq:id3}) instead. This way an arbitrary-order correction is estimated as
\begin{equation}\label{gen_f}
x^{(i)}=\frac{1}{\omega}\mathcal{O}\left((\omega t)^i\right),\quad p^{(i)}=ma_0\mathcal{O}\left((\omega t)^i\right),
\end{equation}
hence the successive terms of the expansion (\ref{eq:basic_exp_x}), (\ref{eq:basic_exp_p}) indeed descend at short times (\ref{eq:t_valid}). 

Now let us turn to calculation of the dependence $\chi_e(t)$ for the time range (\ref{eq:t_valid}). For that it turns out more convenient to re-express $\chi_e^2$ in the form
\begin{equation}
\chi_e^2=-\frac1{m^4}\left(\frac{dp^\mu}{d\tau}\right)^2,
\end{equation}
manifesting $\chi_e$ as electron proper acceleration expressed in Compton units.
Using the expansion (\ref{eq:basic_exp_p}) and taking into account that, due to Eqs.~(\ref{gen_f}) and (\ref{eq:t_valid}), it is enough to track only the leading non-trivial contributions, we obtain
\begin{equation}
\label{eq:chi_sq_1}
\begin{split}
\chi_e^2\approx &
-\frac1{m^4}\left(\frac{dp^{(1)}}{d\tau}\right)^2-\frac2{m^4}\frac{dp^{(1)}}{d\tau}\frac{dp^{(2)}}{d\tau} \\
&-\frac1{m^4}\left[\left(\frac{dp^{(2)}}{d\tau}\right)^2+2\frac{dp^{(1)}}{d\tau}\frac{dp^{(3)}}{d\tau}\right].
\end{split}
\end{equation}
The first (lowest order) term on the RHS 
obviously represents $\chi_e^2(0)$. It formally looks as $\mathcal{O}(t^2)$ but due to cancellations is actually $\mathcal{O}(1)$. The second [formally $\mathcal{O}(t^3)$-order] term vanishes in our ultrarelativistic approximation\footnote{Strictly speaking, by this argument only the leading $\mathcal{O}(t^3)$-order terms should cancel, we come back to this below and in Appendix~\ref{sec:app_rot}.} due to (\ref{eq:ultrarel}), (\ref{eq:dp2}), and (\ref{eq:id2}).
Finally, by noticing that the matrix $F^{(0)}$ is antisymmetric, $dp^{(1)}/d\tau\approx (e\epsilon/m) p^{(1)}$, $dp^{(3)}/d\tau\approx (3e\epsilon/m) p^{(3)}$, and by applying Eqs.~(\ref{eq:dp2}), (\ref{eq:id3}), the last [$\mathcal{O}(t^4)$ order] term in the RHS of Eq.~(\ref{eq:chi_sq_1}) can be cast to 
\begin{equation}
\nonumber
\begin{split}
-\frac{e^2\epsilon^2}{m^6}\left(p^{(2)}\right)^2=\frac{e^2\epsilon^2}{m^6}
p^{(1)}F^{(1)}\frac1{4\epsilon^2-\left(F^{(0)}\right)^2}F^{(1)}p^{(1)}.
\end{split}
\end{equation}

Hence, our final expression for time dependence of the parameter (\ref{eq:chi}) in an arbitrary field takes the form
\begin{equation}
\label{eq:chi_t}
\chi_e^2(t)\approx \chi_e^2(0)+\left(\frac{e^2\epsilon^2\omega_{\text{eff}}}{m^3}\right)^2t^4,
\end{equation}
where
\begin{equation}\label{eq:w_eff}
\begin{split} \omega^2_{\text{eff}}t^4=&x_{\mu}^{(1)}F^\mu_{\;\;\;\nu,\sigma}x^{(1)\sigma}\\
&\times \left(\frac1{4\epsilon^2-\left(F^{(0)}\right)^2}\right)^{\nu}_{\;\;\;\lambda} F^\lambda_{\;\;\;\varkappa,\rho}x^{(1)\rho} x^{(1)\varkappa},
\end{split}
\end{equation}
and $x^{(1)\mu}=f_1^{\mu}t$. Note that with account to Eq.~(\ref{eq:w_eff}) the resulting  Eq.~(\ref{eq:chi_t}) is manifestly Lorentz-invariant.

Formulas (\ref{eq:ultrarel_p}) and (\ref{eq:chi_t}) establish the general short-term behavior of the electron energy and the dynamical quantum parameter $\chi_e(t)$ in an arbitrary field and represent one of the main results of the paper. We apply them below to develop a qualitative theory of an A-type cascade onset in an arbitrary EM field. 

Let us conclude the Section with a few brief remarks. In effect, on the timescale \eqref{eq:t_valid} the particle energy given in Eq.~\eqref{eq:ultrarel_p} is independent of the initial condition, but $\chi_e(t)$ depends on $\chi_e(0)$ in general [see Eq.~\eqref{eq:chi_t}]. If the term $\chi_e^2(0)$ is ignored, then the expression \eqref{eq:chi_t} is simplified further,
\begin{equation}
\label{eq:chi_t_short}
\chi_e(t)\approx \frac{e^2\epsilon^2\omega_{\text{eff}}}{m^3}t^2.
\end{equation}
In particular, for an initially resting seed electron characterized by $\chi_e(0)=E/E_S\ll 1$ [see Eq.~(\ref{eq:chi})], one can easily verify that Eqs.~\eqref{eq:ultrarel_p} and \eqref{eq:chi_t_short} generalize the previously considered cases of a uniformly rotating electric field \cite{fedotov2010limitations,elkina2011qed} (where the effective frequency is $\omega_{\text{eff}}=\omega/2$), linearly and circularly polarized standing waves \cite{bashmakov2014effect}~\footnote{Due to a misprint in \cite{bashmakov2014effect}, we reproduce Eq.~(A16) therein only apart from the superfluous numerical factor 2.}, and multiple colliding beams with magnetic field vanishing at the center of the focus \cite{gelfer2015optimized}. 

However, it follows from \eqref{eq:chi_t} that the transition to \eqref{eq:chi_t_short} is only possible for
\begin{equation}
\label{eq:newcond}
t\gg \frac{m}{e\epsilon}\sqrt{\frac{m\chi_e(0)}{\omega_{\text{eff}}}}.
\end{equation}
Obviously, Eq.~\eqref{eq:newcond} at the same time estimates the timescale on which the dynamical quantum parameter substantially exceeds its initial value. For $a_0\gg 1$ the new restriction (\ref{eq:newcond}) is stronger than the left inequality in Eq.~(\ref{eq:t_valid}), meaning that \eqref{eq:chi_t_short} can never be valid on the whole range \eqref{eq:t_valid}. This is an important refinement over the previous works. On the other hand, by a crude estimate $\epsilon\simeq E$, $\omega_{\text{eff}}\simeq \omega$ and $\chi_e(0)\simeq (E/E_S)\cdot(p_{0\perp}/m)$, the condition \eqref{eq:newcond} is consistent with the right inequality in Eq.~(\ref{eq:t_valid}) if the component of the initial momentum transverse to the field obeys.
\begin{equation}\label{eq:tr_mom}
p_{0\perp}\ll m a_0. 
\end{equation}
This naturally means that the initial transverse momentum is so small that the transverse electron motion is governed by the field. To illustrate these considerations, we study the evolution of $\chi(t)$ explicitly  for a non-resting seed electron in a uniformly rotating electric field in Appendix~\ref{sec:app_rot}. In particular, it is demonstrated there that even though our derivation based on ultrarelativistic approximation can reproduce only the leading contributions in Eq.~\eqref{eq:chi_t}, the subleading ones are indeed negligible under imposing the additional condition Eq.~\eqref{eq:newcond}. Moreover, in spite of possible actual presence of such additional subdominant terms, our Eq.~(\ref{eq:chi_t}) that misses them nevertheless works fine even in the initial range Eq.~\eqref{eq:t_valid} by correctly interpolating between the initial condition and the behavior Eq.~\eqref{eq:chi_t_short} on longer time specified in Eq.~\eqref{eq:newcond}.

According to \eqref{eq:chi_t}, the dynamical quantum parameter of an initially slow particle in an arbitrary nonuniform EM field grows with $t$ on a timescale (\ref{eq:t_valid}), unless the field invariant $\epsilon$ is either strictly zero or  anomalously small. This exceptional situation takes place either for the fields of magnetic type (${\cal F}<0$, ${\cal G}=0$), or for a field close to a plane wave (for which ${\cal F}={\cal G}=0$). In the former case there is always a reference frame in which the electric field locally vanishes and the particle is orbiting around the direction of magnetic field. Obviously, in such a case there is no net acceleration at all. The latter case (or more precisely the paradigmatic case of a weakly focused field) is analyzed in Appendix~\ref{sec:app_sp}. However, for a generic (e.g. tightly focused) field both field invariants are in general substantially non-zero in certain regions, hence slow particles are ultimately accelerated there by the field.

\subsection{Threshold condition for A-type cascade onset in generic field}
\label{sec:criteria}
Selfsustained (A-type) cascades are seeded by slow ($\chi_e(0)\ll 1$) electrons, with the required amount of energy provided entirely due to their ongoing acceleration in the field. In order to launch such a cascade, in the course of acceleration the dynamical quantum parameter of electron should attain the values $\simeq 1$, since otherwise (as long as $\chi_e\ll 1$) the emitted photons are so soft ($\chi_\gamma\sim \chi_e^2\ll \chi_e$) that, according to Eq.~(\ref{eq:W_cr}), their pair photoproduction capability is exponentially suppressed. When such a slow seed electron gets accelerated by a generic field of electric type ($E>H$), its dynamical quantum parameter $\chi_e(t)$ is growing on the time scale (\ref{eq:t_valid}) according to Eq.~(\ref{eq:chi_t}). 

The time spent on average until the event of hard photon emission is estimated by \footnote{More precisely by $\int_0^{t_\mathrm{free}} W_\mathrm{rad}(t)\,dt\simeq 1$, which is the same up to a numerical factor $\simeq 1$.} $t_\mathrm{free}\simeq 1/W_\mathrm{rad}$. The probability rate $W_\mathrm{rad}(t)$ increases along with the electron dynamical quantum parameter $\chi_e(t)$ in the course of acceleration. Hence, by picking up the asymptotic in Eq.~(\ref{eq:W_rad}) corresponding to $\chi_e\gg 1$  (we assume that it is lingered over till $\chi_e\gtrsim 1$) and substituting (\ref{eq:ultrarel_p}), 
we obtain an estimate
\begin{equation}
\label{eq:chi_mean}
\chi_e(t_\mathrm{free})\simeq \mu^{3/2},
\end{equation}
where, for brevity, we have introduced the dimensionless field strength parameter $\mu=\epsilon/\alpha E_S$. Assuming $\chi_e(t_\mathrm{free})\gg\chi_e(0)$ [equivalently, if $t_\mathrm{free}$ matches the condition \eqref{eq:newcond}], one can express $t_\mathrm{free}$ explicitly using Eq.~ \eqref{eq:chi_t_short}:
\begin{equation}
\label{eq:t_free}
t_\mathrm{free}\simeq  \frac{1}{\varkappa\mu^{1/4}\omega_\mathrm{eff}}.
\end{equation}
Here we introduced the numerical coefficient $\varkappa=\sqrt{\alpha^2m/\omega_\mathrm{eff}}$. For a field carrier frequency in the optical range (assuming $\omega_\mathrm{eff}\sim \omega\simeq 1$ eV) we have $\varkappa\simeq 5$.

According to Eq.~\eqref{eq:chi_mean}, the emitted photons are capable for pair photoproduction [$\chi_e(t_\mathrm{free})\gtrsim 1$, as implied by Eq.~\eqref{eq:W_cr}] if $\mu\gtrsim 1$, or
\begin{equation}
\label{eq:general_condition}
\epsilon\gtrsim\alpha E_S.
\end{equation}
After emitting a hard photon the electron  slows down, as both its energy and dynamical quantum parameter are partially transferred to the photon. In the regime $\mu\gtrsim 1$ ($\chi_e\gtrsim 1$) their relative loss is of the order of unity. Then we can think of the process as coming back to the initial state and repeating on and on until the electron escapes from the strong field region. As for the emitted photon, it is created with $\chi_\gamma\sim \chi_e$ and $\varepsilon_\gamma\sim\varepsilon_e$, hence produces a pair during about the same\footnote{In this qualitative discussion, we neglect the overall numerical factors in the asymptotics of Eq.~(\ref{eq:W_rad}) and (\ref{eq:W_cr}).} time $t_\mathrm{free}$. 

For the reasons outlined above, we promote Eq.~(\ref{eq:general_condition}) as a criterion for selfsustainability of a cascade. It naturally generalizes the criterion $E\gtrsim \alpha E_S$ \cite{fedotov2010limitations} by (i) replacing the field strength $E$ in a laboratory frame by the electric field invariant $\epsilon$, thus taking into account the effect of the magnetic field; and (ii) replacing the assumption of Ref. \cite{fedotov2010limitations} of initially resting seed electron by a weaker one $\chi_e(0)\lesssim1$. Noteworthy, the component of the electron momentum transverse to the field (no matter before or after hard photon emission) can be estimated as $p_{e\perp}\simeq (m/\alpha)\sqrt{\mu}$, and satisfies the condition Eq.~(\ref{eq:tr_mom}). This confirms the consistency of our approximations.

In order to better explain the meaning of replacing in a generic situation the electric field strength $E$ with the electric field invariant $\epsilon$, let us recall that the field invariants $\epsilon$ and $\eta$  represent nothing but the electric and magnetic field strengths in a special reference frame where they are parallel (it can be thought of as a local `proper frame' of the field, as the Poynting vector vanishes). 
In this frame our criterion (\ref{eq:general_condition}) 
literally coincides with $E\gtrsim \alpha E_S$ suggested previously \cite{fedotov2010limitations}. 
Still, the setup in the `proper frame'  differs from the case of an initially slow seed electron in a uniformly rotating electric field by that (i) the electron, being initially slow in the laboratory frame, now moves transversely to the fields;
and (ii) the magnetic field of strength $\eta$ parallel to the electric field is now present. However, with regard to (i), the Lorentz-invariant condition $\chi_e(t_{\mathrm{free}})\gg\chi_e(0)$ ensures that this transverse initial momentum is insubstantial in the sense of Eq.~(\ref{eq:tr_mom}). As for (ii), since the electron moves deviating only slightly from the common direction of the fields, the resulting Lorentz force is  negligible.

In principle, in order to ensure a substantial cascade multiplicity, one needs to require in addition to Eq.~(\ref{eq:general_condition}) also $t_{\text{free}}\ll t_\mathrm{esc}$, where $t_\mathrm{esc}$ is the time of escape of the particles from the strong field region. However, it is not easy to give a reasonable estimate of the escape time $t_\mathrm{esc}$ in general, since the cascade structure and long-time behavior are rather intricate, with possible additional complications imposed by radiative trapping \cite{elkina2014improving, fedotov2014radiation}. On the other hand, since anyway $t_{\mathrm{esc}}\gtrsim \pi/\omega$, the right inequality in Eq.~\eqref{eq:t_valid} is more restrictive. Noteworthy, in Ref. \cite{fedotov2010limitations} it was assumed that for a laser field focused to a diffractive limit $t_\mathrm{esc}\simeq \lambda/2=\pi/\omega$. In that context, the condition $t_{\text{free}}\ll t_\mathrm{esc}$ appeared to be weaker than Eq.~\eqref{eq:general_condition}. As we will see further, this may not be always the case in general. 

Finally, note that the cascade onset threshold clearly depends on its precise  definition. For the reasons discussed in Ref.~\cite{fedotov2016threshold}, the criterion \eqref{eq:general_condition} might overestimate the actual thresholds observed in  particular numerical simulations. Not necessarily precise, Eq.~\eqref{eq:general_condition} is useful as a universal and transparent guiding mark for a wide class of external field models. 

\section{Cascades in a single focused laser pulse}
\label{sec:single_pulse}
\subsection{Onset of a selfsustained cascade}
\label{sec:sp_estimates}
To illustrate general considerations, consider an A-type cascade onset in a single focused laser pulse. Let us derive the corresponding threshold condition explicitly in terms of the field parameters. To describe the laser field, we use a model of a monochromatic focused circularly $e$-polarized Gaussian beam proposed in Refs. \cite{narozhny2000scattering, narozhny2004e+}. It is parametrized by the peak EM field strength at the focus $E_0$, the angular aperture $\Delta$, which is assumed to be small $\Delta\ll 1$, and the frequency $\omega$. The focal spot radius and the Rayleigh length of the laser beam are given by $R=1/\omega\Delta$ and $L=R/\Delta$, respectively. The diffraction limit is reached for $\Delta\sim 0.3$. We assume the laser beam propagates along $z$-axis, and that the focal center coincides with the origin $\mathbf{r}=0$. The expressions for the EM field are given in Appendix~\ref{sec:app_sp}, see Eqs.~\eqref{app:E_e}, \eqref{app:H_e}.

Suppose a seed electron is placed in the focal region of the laser beam\footnote{The discussion of injecting seed particles into the focus is postponed to Section~\ref{sec:collapse_and_revival}.}. For the sake of simplicity, we assume that the electron is initially located precisely at the center of the focus with $\chi_e(0)\ll 1$ [e.g. $\chi_e(0)=E_0/E_S\ll 1$ if the electron is initially at rest]. In the vicinity of this point, the EM field under consideration is of electric type [see Eq.~\eqref{app:F0}], therefore the electron is accelerated. On the timescale (\ref{eq:t_valid}) its energy and the parameter $\chi_e$ can be approximated by Eqs.~\eqref{eq:ultrarel_p} and \eqref{eq:chi_t_short}, respectively. By expanding the field near the initial position $\vec{r}=0$ of the electron, we obtain $\epsilon\approx 2\sqrt{2}\Delta E_0$ and $\omega_\mathrm{eff}=17\sqrt{2}\Delta^3\omega$ (see Appendix~\ref{sec:app_sp} for details). This yields the approximate expressions 
\begin{gather}
\label{eq:main1}
\varepsilon(t)\approx 2\sqrt{2}\Delta e E_0t,\\
\label{eq:main2}
\chi_e(t)\approx 8\Delta^5\tilde{\varkappa}\left(\frac{E_0}{\alpha E_S}\right)^2(\omega t)^2,
\end{gather}
which are valid on the timescale
\begin{equation}
\label{eq:main_t}
\frac{\sqrt{\chi_e(0)}}{\tilde{\varkappa} \mu \Delta^{3/2}}\ll \omega t\ll \pi
\end{equation}
[recall that the left inequality in Eq.~\eqref{eq:t_valid} needs to be superseded with a stronger one Eq.~\eqref{eq:newcond}].  
Here $\mu=\epsilon/\alpha E_S\approx 2\sqrt{2}\Delta E_0/\alpha E_S$, and the constant $\tilde{\varkappa}=(17\sqrt{2}\alpha^2m/\omega)^{1/2}\simeq 25$ for $\omega = 1$ eV.
Noteworthy, the inequality between the outermost terms in Eq.~\eqref{eq:main_t} restricts the laser beam angular aperture  from below:
\begin{equation}
\label{eq:acc_cap}
\Delta\gg a_0^{-1/5}
\end{equation}
(from now on we switch to notation $a_0=eE_0/m\omega$). 

\begin{figure}
	\includegraphics[width=0.46\textwidth]{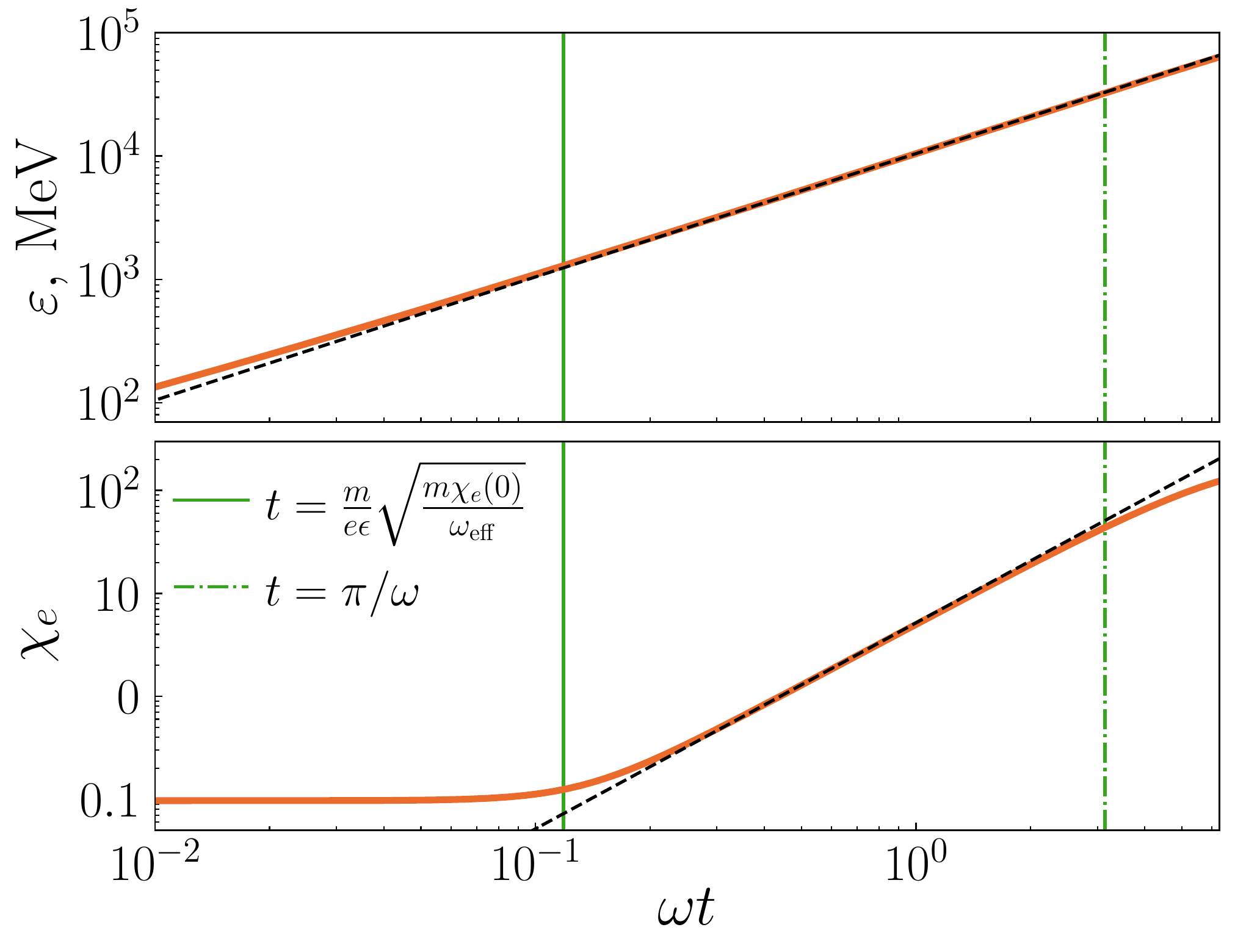}
	\caption{Time dependence of energy $\varepsilon$ and the parameter $\chi_e$ of an electron in a laser beam given in Eqs.~\eqref{app:E_e}, \eqref{app:H_e}: the approximate expressions given in Eqs.~\eqref{eq:main1}, \eqref{eq:main2} (dashed line) vs simulation (orange solid line). The green solid and dot-dashed vertical lines indicate the endpoints of the interval of validity of our approximation [see Eq.~\eqref{eq:main_t}]. The electron is initially at rest at the center of the focus ($\vec{r}=0$); the laser beam parameters are $\omega=1$ eV, $E_0=0.1E_S$ ($a_0\approx5\times 10^4$), $\Delta=0.1$.}
	\label{fig:main_check}
\end{figure}

A test of our approximate expressions \eqref{eq:main1}, \eqref{eq:main2} against the numerical simulation of the electron motion in the laser field is displayed in Fig.~\ref{fig:main_check}. Clearly, they are in good agreement inside the interval of validity \eqref{eq:main_t}. 

Next, we estimate the mean free path time $t_{\text{free}}$ of the electron with respect to photon emission. According to Eq.~\eqref{eq:t_free}, we obtain
\begin{equation}
\label{eq:t_free_SP}
\omega t_{\text{free}}\simeq \frac 1 {\tilde{\varkappa} \mu^{1/4} \Delta^{3/2}}.
\end{equation}
Provided that the criterion $\mu\gtrsim 1$ is satisfied [see Eq.~\eqref{eq:general_condition}], which in this case reads
\begin{equation}
\label{eq:single_pulse_cond1}
\frac{E_0}{E_S}\gtrsim \frac{\alpha}{2\sqrt{2}\Delta},
\end{equation}
there is a substantial probability that the emitted photon is hard (as discussed in Sec.~\ref{sec:criteria}). Noteworthy, unlike for the uniform rotating electric field (cf. \cite{fedotov2010limitations}), this condition incorporates both the field strength $E_0$ and the angular aperture $\Delta$. In the plane wave limit $\Delta\to 0$ the threshold value naturally tends to infinity.

The mean free path time $t_{\text{free}}$ should match the approximation validity conditions given in Eq.~\eqref{eq:main_t}. This yields additional constraints on the field parameters. 
The inequality in the LHS simply means that $\chi_e(t_{\text{free}})=\mu^{3/2}\gg \chi_e(0)$ [see Eqs.~\eqref{eq:chi_t},\eqref{eq:newcond}]. 
For an initially slow electron this condition follows automatically from Eq.~\eqref{eq:single_pulse_cond1}. As for the inequality in the RHS of Eq.~\eqref{eq:main_t}, by plugging Eq.~\eqref{eq:t_free_SP} we arrive at:
\begin{equation}
\label{eq:single_pulse_cond2}
\frac{E_0}{E_S}\gtrsim \frac{\alpha}{2\sqrt{2}\pi^4\tilde{\varkappa}^4\Delta^7}.
\end{equation}
Note that it is stronger than Eq.~\eqref{eq:single_pulse_cond1} for weak focusing $\Delta< (\pi^2\tilde{\varkappa})^{-1/3}$ ($\Delta\lesssim 0.05$ for $\omega=1$ eV).

Let us also estimate the time $t_\mathrm{esc}$ needed for the electron to escape from the strong field region. As already mentioned, it is hard to follow the long-time cascade dynamics, which is in general rather intricate, at best we can estimate $t_\mathrm{esc}$ from below.
Recall that over a short time the electron position evolves as $x^{\mu(1)}=f_1^\mu t$ [see Eq.~(\ref{eq:ultrarel1})]. In the  case under consideration $f_{1\perp}/f_{1z}\sim\Delta$, where $f_{1\perp}$ and $f_{1z}\sim 1$ are the transverse and longitudinal components of the eigenvector $f_1^\mu$ [see Eq.~(\ref{app:f_i})]. Therefore, by assuming $|f_{1\perp}| t_\mathrm{esc}\sim R$, we obtain an estimate $t_{\text{esc}}\simeq R/\Delta=1/\omega\Delta^2=L$. As $t_{\text{esc}}\gg \pi/\omega$, the consecutive processes of acceleration and hard photon emission can repeat multiple times before the particles can escape the focal region. 
Thus promoting Eqs.~ \eqref{eq:single_pulse_cond1} and \eqref{eq:single_pulse_cond2} as the criteria of an A-type cascade onset in the field of a single focused laser beam looks reasonable.

\begin{figure}
	\includegraphics[width=0.49\textwidth]{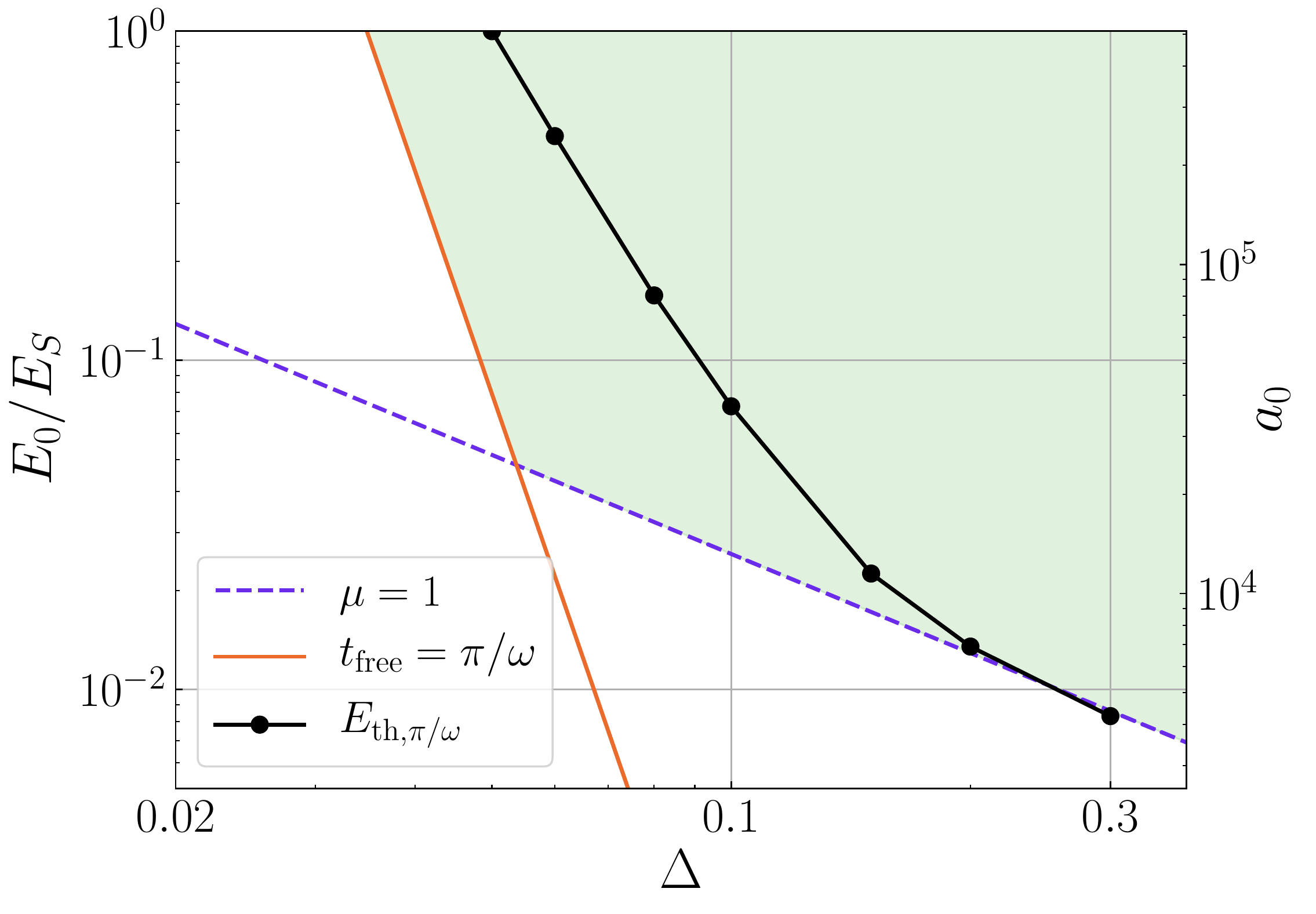}
	\caption{Numerically calculated threshold field strength $E_{\text{th}}$, required to initiate an A-type cascade with a seed electron, initially at rest at $\mathbf{r}=0$, in dependence on the laser beam angular aperture $\Delta$  (solid line with circles). The threshold is defined by that a single $e^-e^+$ pair is created per initial $e^-$ on average during the laser half-period $\pi/\omega$. Inside the green area the criteria given in Eq.~\eqref{eq:single_pulse_cond1} (bounded by the dashed line) and Eq.~\eqref{eq:single_pulse_cond2} (bounded by the solid line) are fulfilled simultaneously.}
	\label{fig:cascade_init}
\end{figure}

To test the above conclusions, we performed a numerical simulation using the Monte-Carlo code described in Ref.~\cite{mironov2016generation}. We consider basically the same laser field model as in the discussion above, only modified by including a Gaussian-shaped temporal envelope the same way as in Ref.~\cite{narozhny2000scattering}. We set the field frequency $\omega=1$ eV, the laser pulse duration $\tau_L=10$ fs and varied the peak field strength $E_0$ and the angular aperture $\Delta$ through the simulation runs. In each  simulation, a single seed electron was placed at rest at the center of the focus at the moment when the temporal envelope achieves maximum (i.e. the electron experiences the peak field strength). The results were averaged over an ensemble of $10^3$ identical initial particles for each parameter setting. All simulations were carried out with the values of the parameters close to the expected cascade onset threshold. During the whole simulation time we operated at low particle densities, therefore it was legitimate to neglect all the plasma and laser absorption effects, as discussed, e.g., in Refs. \cite{nerush2011laser, ridgers2013dense, grismayer2017seeded}. 

The cascade onset threshold is not sharp and was practically defined akin to Ref.~\cite{gelfer2015optimized}. Namely, we deem a cascade has taken place if at least one $e^-e^+$ pair per single initial $e^-$ is created on average during a half-period $\pi/\omega$ of the laser field. For given values of $\Delta$, we run a series of simulations with variable laser peak field strength $E_0$ and analyze the simulation data set to identify the corresponding threshold value $E_{\text{th}}$.

The results are presented in Fig.~\ref{fig:cascade_init}. One can see that the simulation results for the threshold field parameters are in reasonable agreement with the criteria given in Eqs.~\eqref{eq:single_pulse_cond1} and \eqref{eq:single_pulse_cond2}. Furthermore, for stronger focusing (higher values of $\Delta$) the dependence $E_{\text{th}}(\Delta)$ is well described by Eq.~\eqref{eq:single_pulse_cond1}, while for weaker one (lower values of $\Delta$) it tends to the line corresponding to Eq.~\eqref{eq:single_pulse_cond2}.
As already mentioned, a precise threshold value depends on the adopted  duration for electron doubling. For example, by running simulations for $\Delta=0.1$ over half duration of the laser pulse instead of half-period, on average a pair is created per electron by the end of a simulation for $E_{\text{th},\tau_L/2}\approx 0.03 E_S$ instead of $E_{\text{th},\pi/\omega}\approx 0.07 E_S$. While quantitative agreement is peculiar to our particular definition of the threshold value (in a sense just justifying its reasonability), the agreement of the simulation data distribution with the slopes of the lines corresponding to Eqs.~\eqref{eq:single_pulse_cond1} and \eqref{eq:single_pulse_cond2} is more important and demonstrates the qualitative rationality of our analysis and of the proposed criteria in general.

Note that according to the simulations, for $E_0=E_S$ (the corresponding intensity $I_S\sim 10^{29}$~W/cm$^2$) an A-type cascade is initiated if $\Delta\gtrsim 0.05$. On the other hand, one needs a substantially lower laser intensity to initiate such a cascade with a stronger focused pulse, e.g. for $\Delta=0.15$ one needs $I=E_0^2/4\pi\sim10^{26}$ W/cm$^2$. This means that it is possible to control the multiplicity of a cascade by varying $\Delta$. In particular, decreasing $\Delta$ suppresses the cascade multiplicity, thus allowing to attain higher field strength without facing a depletion of the laser pulse (cf. Refs.~\cite{fedotov2010limitations, nerush2011laser, grismayer2017seeded, tamburini2017laser}).

\subsection{Collision with GeV electrons}
\label{sec:collapse_and_revival}
In reality, seed particles can be delivered to the laser focus by colliding a laser pulse head-on with a bunch of high-energy electrons. If their energy is high enough, then an S-type cascade develops on impact \cite{sokolov2010pair,bulanov2013electromagnetic,blackburn2017scaling,vranic2018multi,magnusson2019multiple,wan2020ultrarelativistic}. Suppose a photon emitted at a mid-stage produces a slow electron or positron at the central region of the focus. We call a particle ``slow'' if its motion is driven essentially by the laser field. If the laser field parameters satisfy  Eqs.~\eqref{eq:single_pulse_cond1} and \eqref{eq:single_pulse_cond2}, then such a slow particle can further seed an A-type cascade. Such a cascade transformation was studied previously in a different setup with two counterpropagating laser pulses, see Refs.~\cite{mironov2014collapse,mironov2016generation,mironov2017observable}. 

\begin{figure}
	\hspace*{-4.3mm}
	\includegraphics[width=0.467\textwidth]{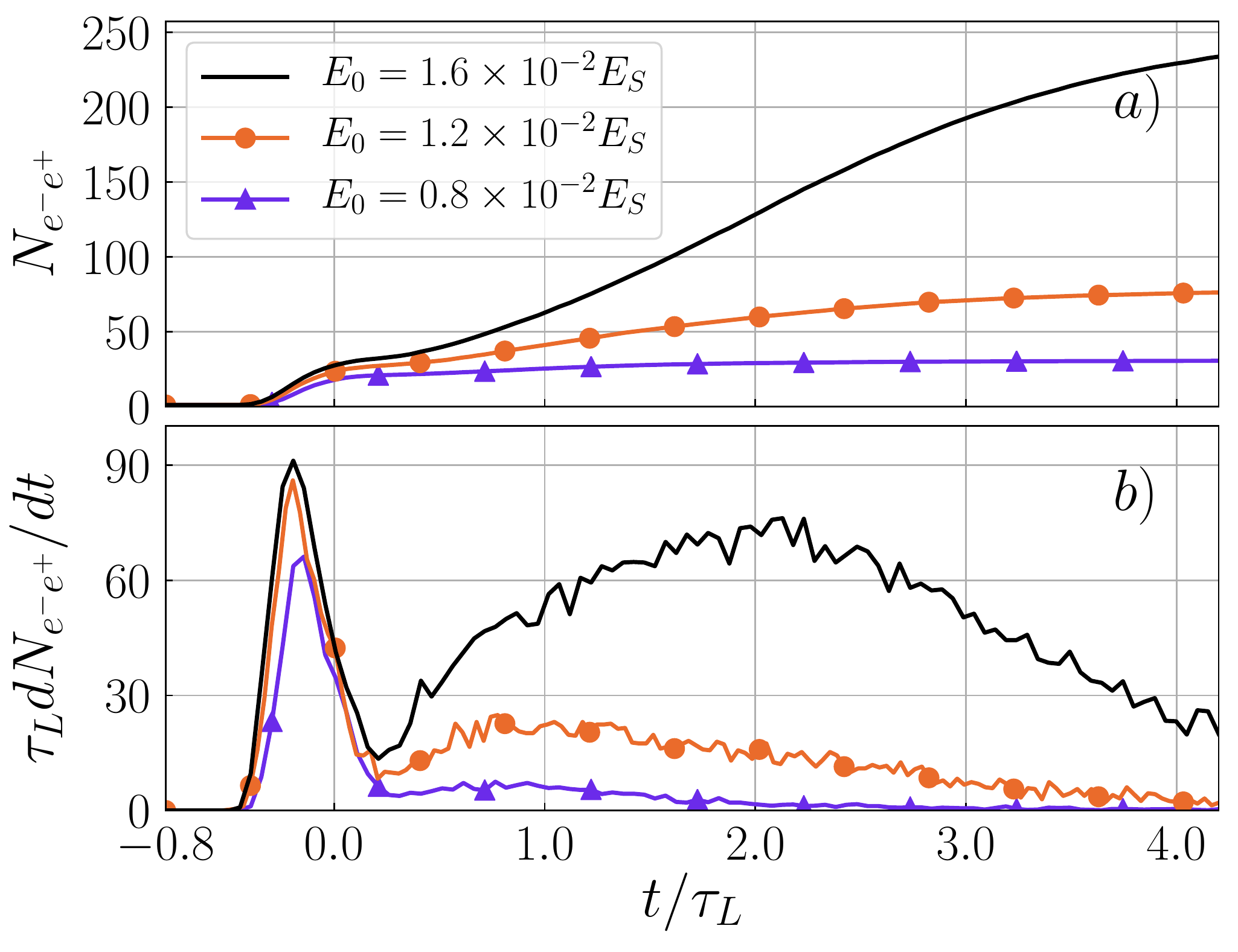}
	\caption{The time dependence of the number $N_{e^-e^+}$ of pairs (a) and the pair creation rate $dN_{e^-e^+}/dt$ (b) for different values of the peak laser field strength $E_0$ (the corresponding values of $a_0=eE_0/m\omega$: $4.1\times 10^3$, $6.1\times 10^3$ and $8.2\times 10^3$ and of the intensities $I=E_0^2/4\pi$ [W/cm$^2$]: $3.0\times 10^{25}$, $6.7\times10^{25}$ and $1.2\times10^{26}$).}
	\label{fig:N}
\end{figure}

\begin{figure}
	\includegraphics[width=0.45\textwidth]{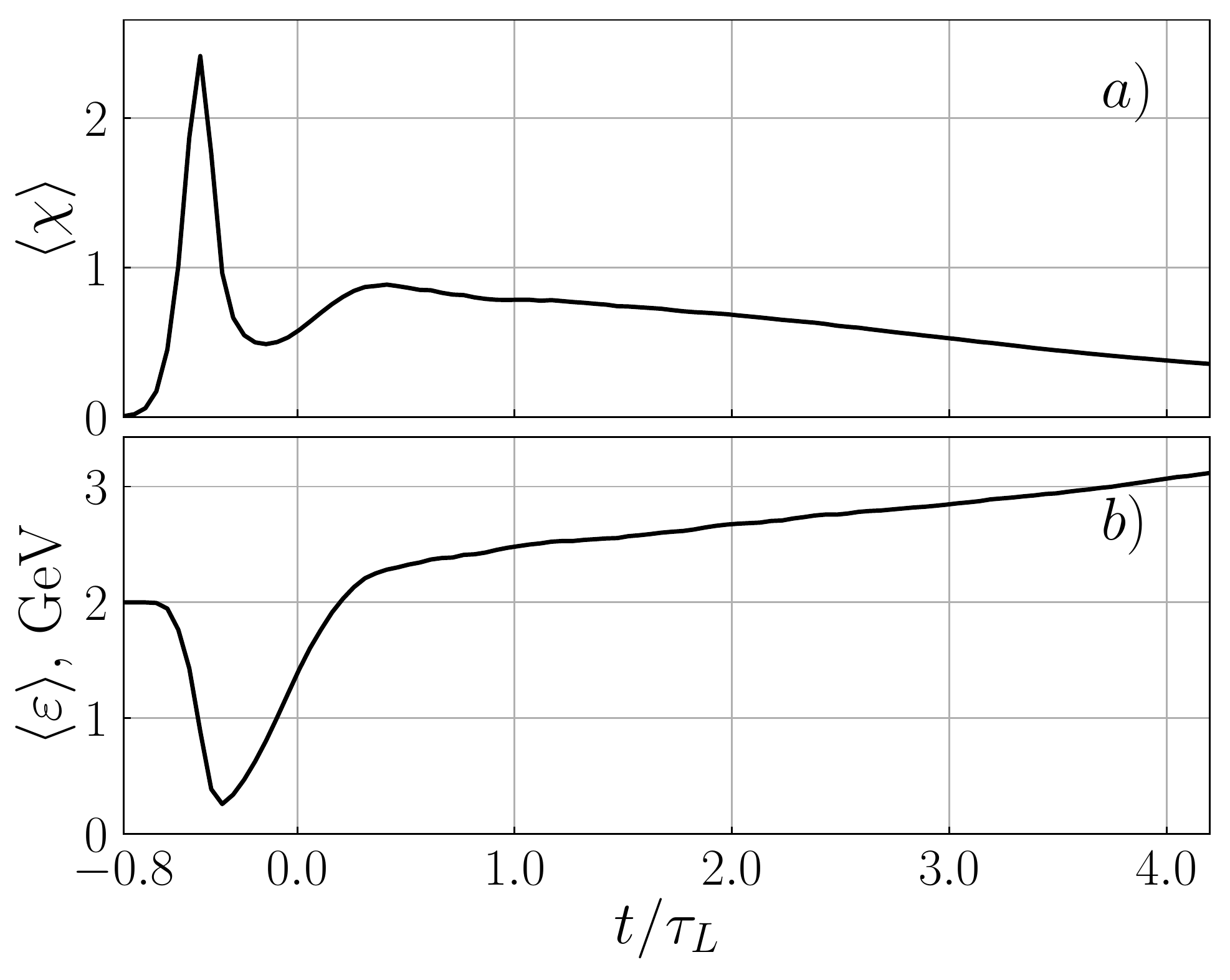}
	\caption{Values of the parameter $\chi$ (a) and energy $\varepsilon$ (b), averaged over electrons in the cascade versus time for $E_0=1.6\times 10^{-2}E_S$ (the corresponding$a_0=8.2\times 10^3$ and $I=1.2\times10^{26}$ W/cm$^2$).}
	\label{fig:chi_energy}
\end{figure}

\begin{figure*}
	\begin{minipage}{0.45\linewidth}
		\includegraphics[width=\linewidth, trim={12, 11, 8, 5}, clip]{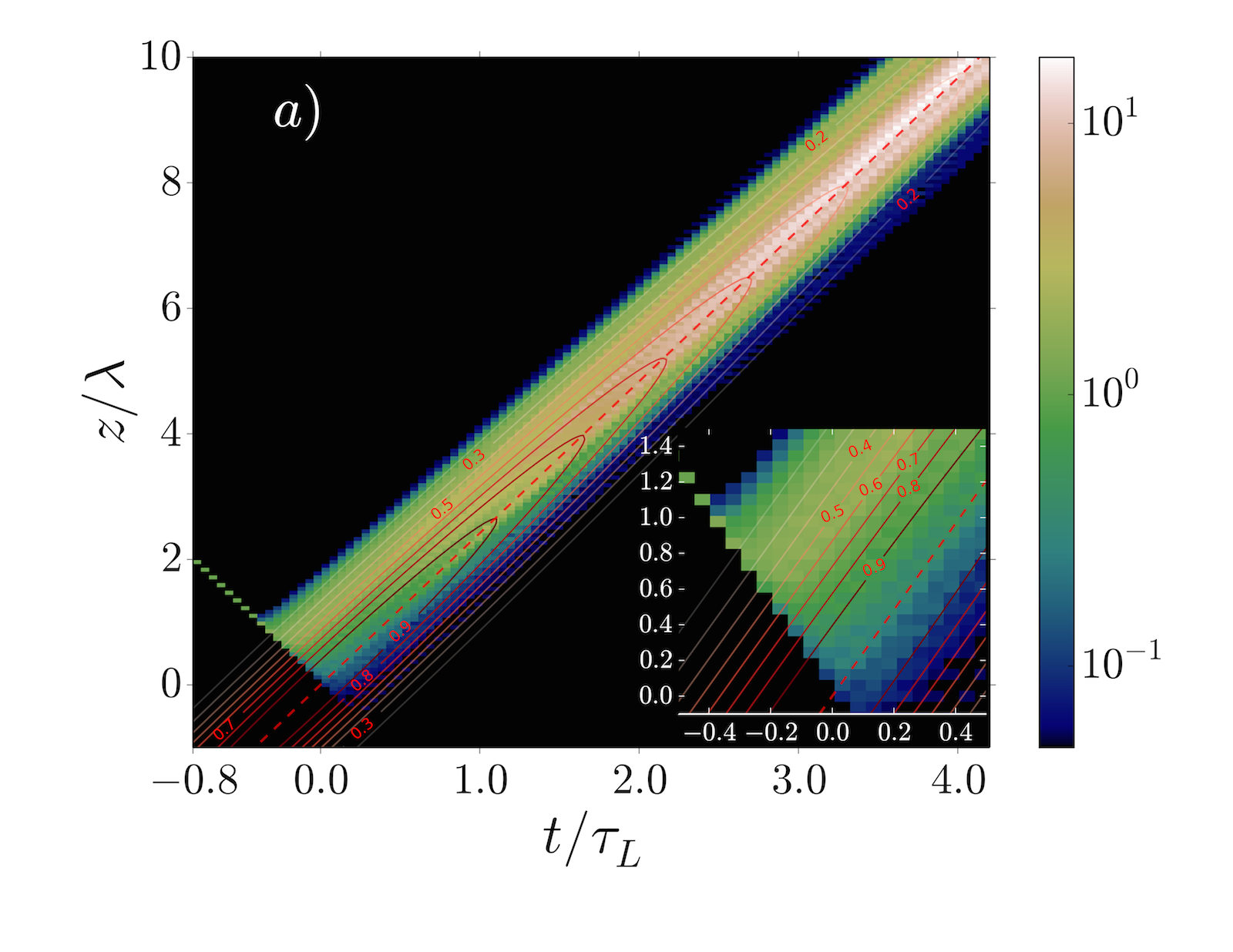}
	\end{minipage}
	\begin{minipage}{0.45\linewidth}
		\includegraphics[width=\linewidth, trim={15, 0, 8, -10}, clip]{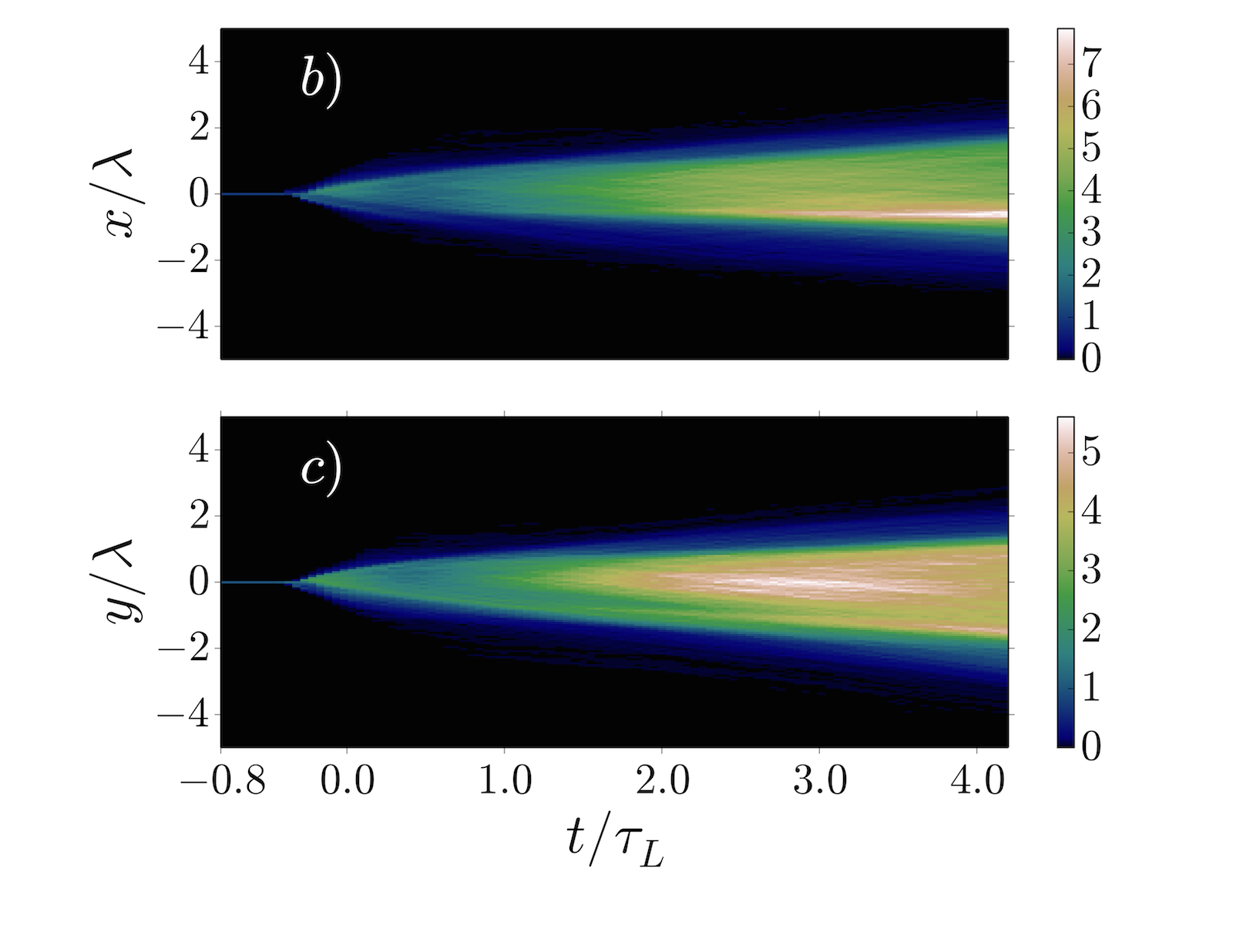}
	\end{minipage}
	\caption{Evolution of the electron density: a) $d^2N_{e^-}/dtdz$ (in log scale), contours in red indicate the amplitude of the electric field $E(z,t)/E_0$, red dashed line --- the center of the laser pulse (zoomed at the inset); b) $d^2N_{e^-}/dtdx$ (in a. u.); c) $d^2N_{e^-}/dtdy$ (in a. u.).  Here $E_0=1.6\times 10^{-2}E_S$ ($a_0=8.2\times 10^3$, $I=1.2\times10^{26}$ W/cm$^2$).}
	\label{fig:density_single_pulse}
\end{figure*}

\begin{figure*}
	\begin{minipage}{0.31\linewidth}
		\includegraphics[width=\linewidth, trim={25, 25, 22, 10}, clip]{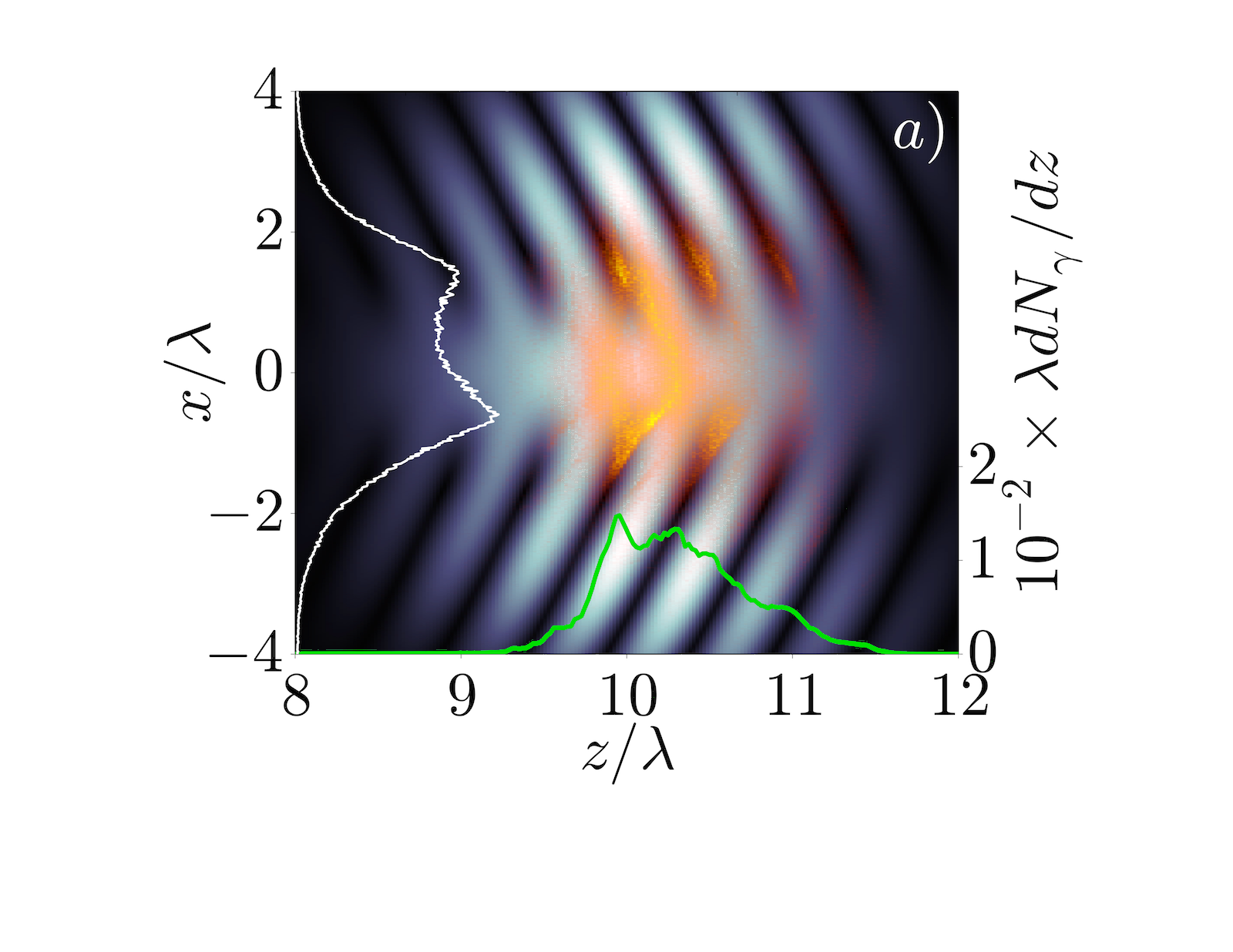}
	\end{minipage}
	\begin{minipage}{0.31\linewidth}
		\includegraphics[width=\linewidth, trim={25, 25, 22, 10}, clip]{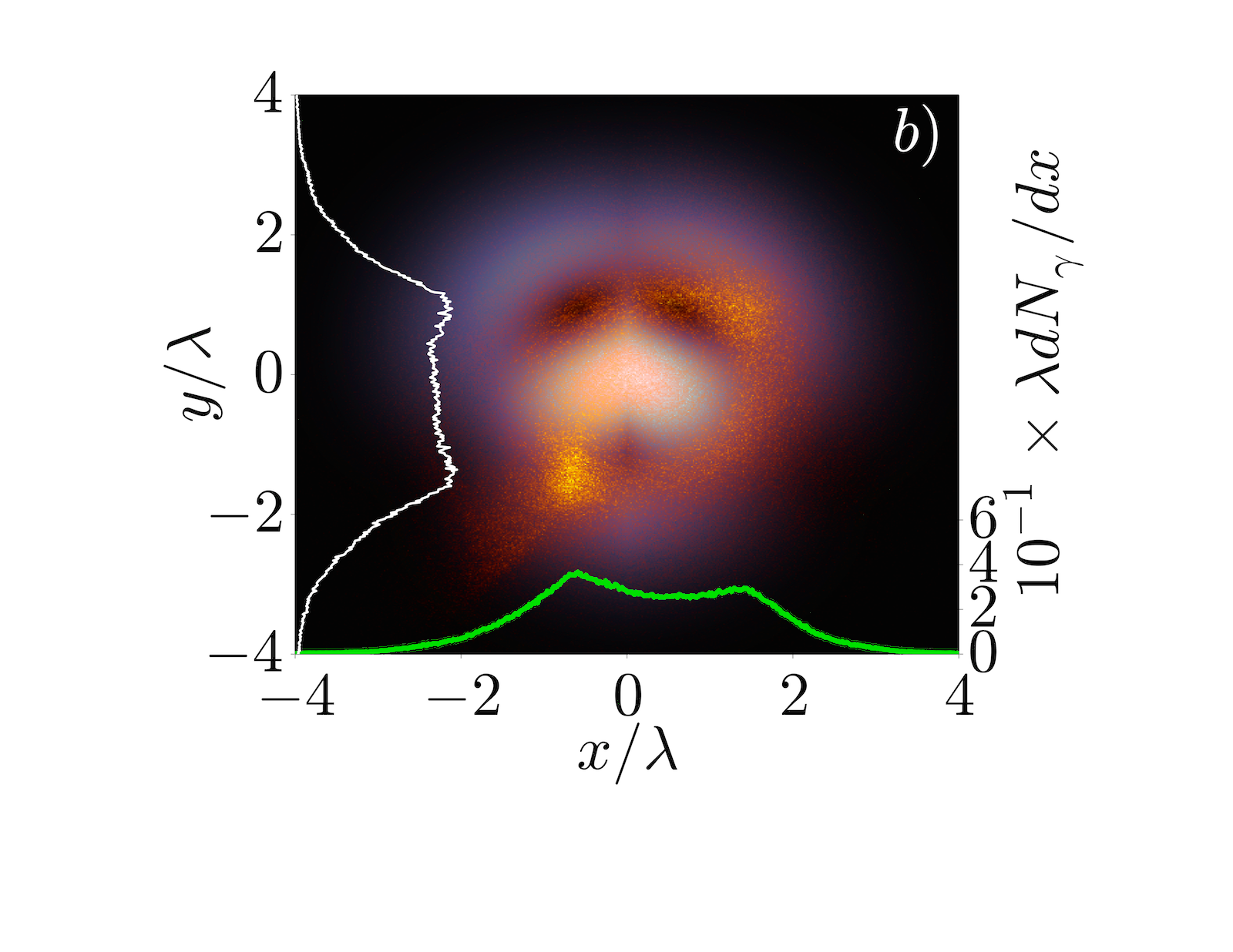}
	\end{minipage}
	\begin{minipage}{0.31\linewidth}
		\includegraphics[width=\linewidth, trim={10, 0, 0, 9}, clip]{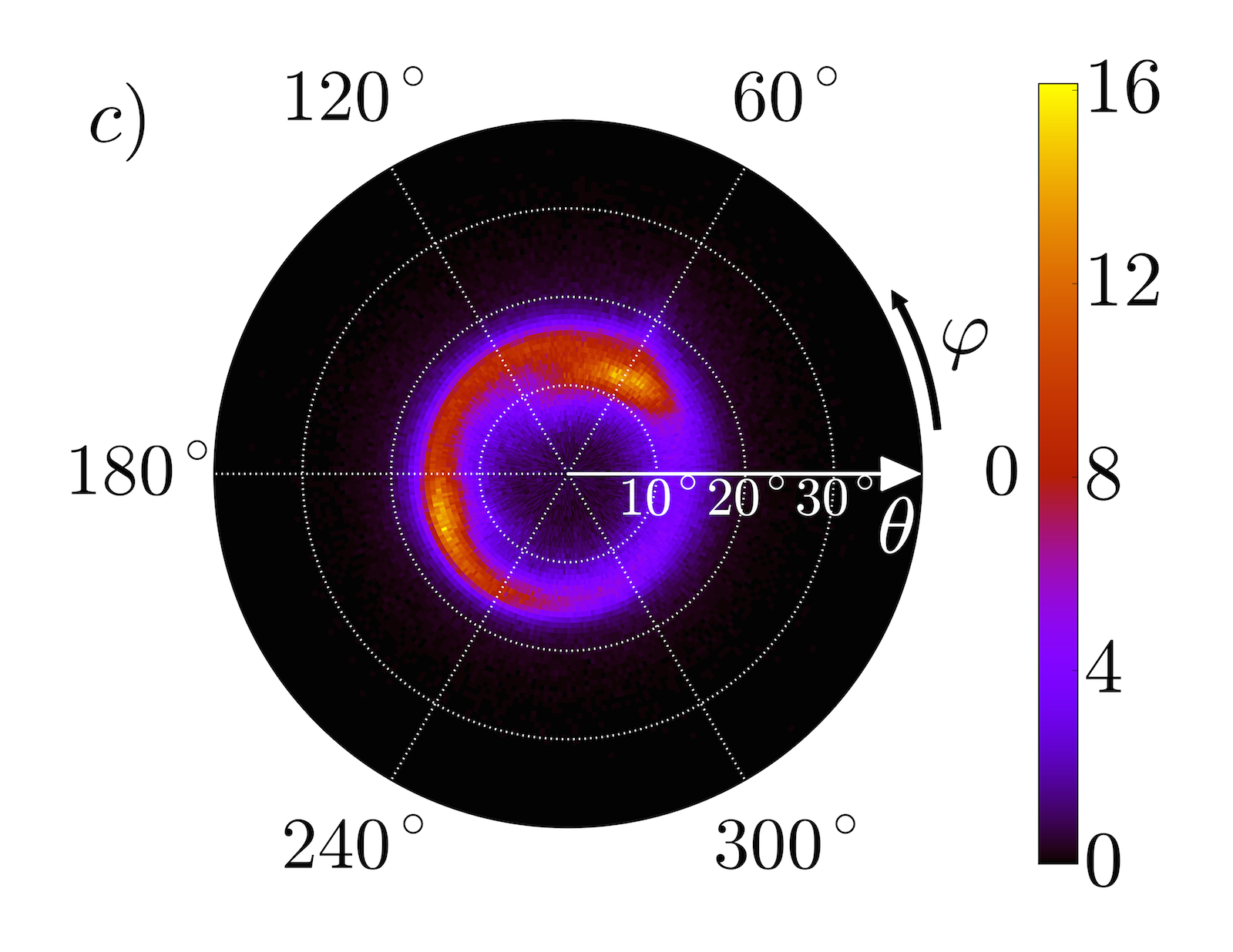}
	\end{minipage}
	\caption{a-b) The spatial distribution of photons (in a. u., in orange tones) $d^2N_\gamma/dzdx$ (a) and $d^2N_\gamma/dxdy$ (b) at time $t=4.2\tau_L$. The monochrome distribution corresponds to the amplitude of the electric field (in a. u.). Lighter color corresponds to higher value. Green solid line: distribution of photons $dN_\gamma/dz$ (a), $dN_\gamma/dx$ (b) (labelled on the right). White solid line: distribution of photons $dN_\gamma/dx$ (a), $dN_\gamma/dy$ (b) (in a. u.) c) The angular distribution of photons $d^2N_\gamma/\sin \theta d\theta d\varphi$ (in a. u.), where $\theta$ is the polar angle between the photon momentum $\mathbf{k}$ and $z$-axis, $\varphi$ is the azimuthal angle between $\mathbf{k}$ and $x$-axis. All the distributions are normalized to the number of initial electrons. Here $E_0=1.6\times 10^{-2}E_S$ ($a_0=8.2\times 10^3$, $I=1.2\times10^{26}$ W/cm$^2$).}
	
	\label{fig:gamma_pulse}
\end{figure*}

\begin{figure}
	\includegraphics[width=0.45\textwidth]{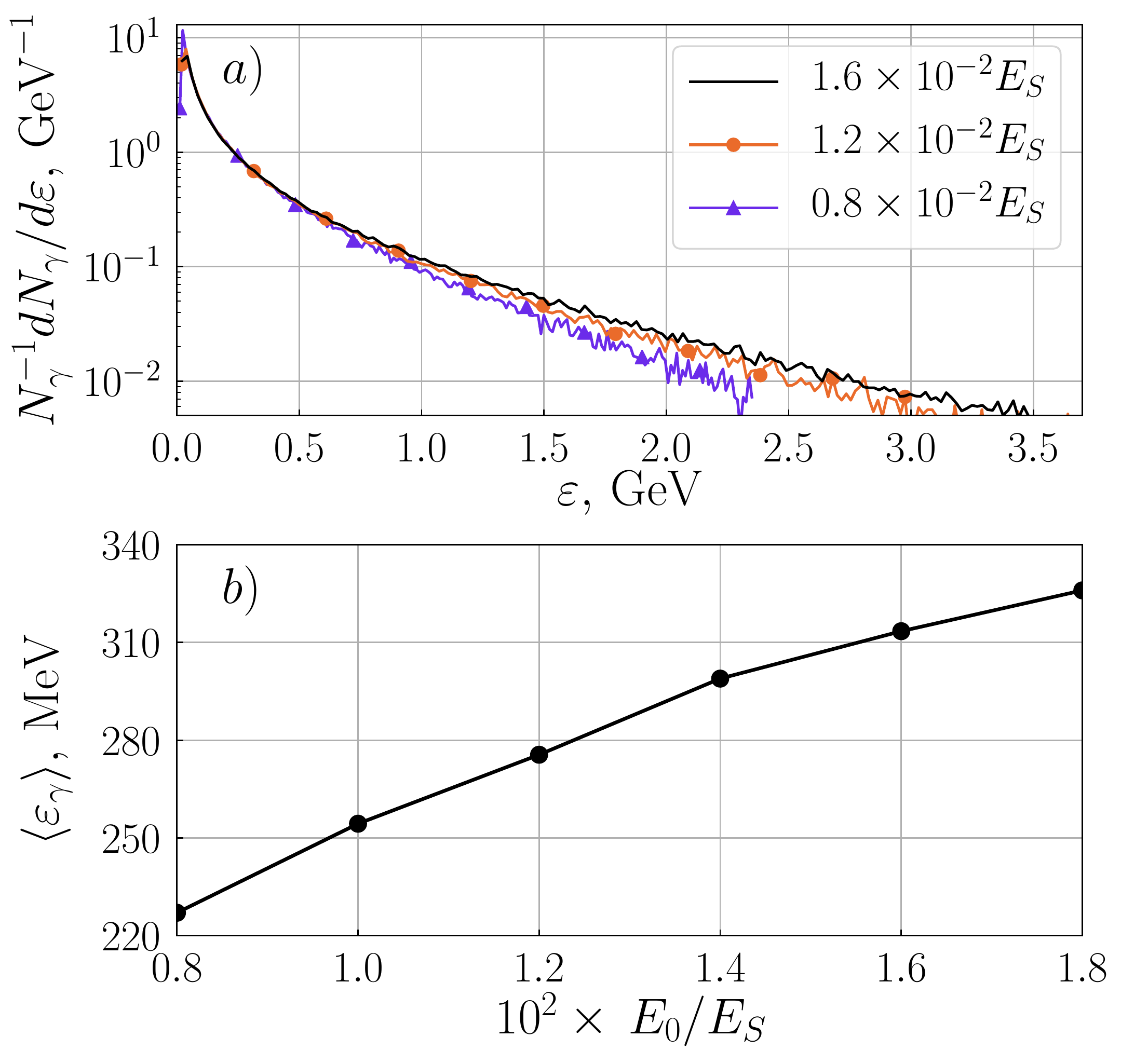}
	\caption{Energy spectra  $N_\gamma^{-1}\frac{dN_\gamma}{d\varepsilon}$ (a) and mean energy $\langle \varepsilon_\gamma \rangle$ (b) of the emitted photons at $t=4.2\tau_L$ for different values of $E_0$.}
	\label{fig:photon_spectra}
\end{figure}

Here we report the results of Monte-Carlo simulations of the cascade dynamics in a single focused laser pulse, assuming that the cascade is seeded by a counterpropagating  GeV electron. Initially, the laser pulse and the electron are set on $z$-axis and propagate in opposite directions. The laser pulse is maximally focused at $t=0$ with the focal spot centered at the origin. The initial location of the electron is such as to reach the same point at $t=0$ in the absence of the laser pulse. In each simulation we start with a single electron of energy $\varepsilon_0=2$ GeV (a close value for $\varepsilon_0$  was used in Ref.~\cite{mironov2014collapse}). The field model, frequency and duration of the pulse are the same as in Sec.~\ref{sec:sp_estimates} ($\omega=1$ eV, $\tau_L=10$ fs). Other laser pulse parameters are tweaked to be close to the threshold value of an A-type cascade onset [see Eqs.~\eqref{eq:single_pulse_cond1}, \eqref{eq:single_pulse_cond2} and Fig.~\ref{fig:cascade_init}]. In particular, we assume that $\Delta=0.1$ and $E_0\sim 10^{-2}E_S$ ($a_0\approx 5\times10^3$). The results of simulations are presented in Figs.~\ref{fig:N}-\ref{fig:photon_spectra}. 

The cascade profiles for three such values of $E_0$ are displayed in Fig.~\ref{fig:N}. For the lowest value $E_0=0.8\times 10^{-2} E_S$ one can see a sharp peak in the pair production rate $dN_{e^-e^+}/dt$. The peak forms when electron is passing a front wing of the pulse. But for higher values of $E_0$ this peak is followed by a long hump, so that the cascade evolution can be naturally divided into two stages. Note that whereas the number of pairs $N_{e^-e^+}$ created during the first stage is about the same for all the values of $E_0$, at the second stage $N_{e^-e^+}$ grows substantially with $E_0$. This feature (see Ref.~\cite{mironov2017observable}) suggests to associate the first peak of the production rate with an S-type cascade and the follow-up hump with an A-type cascade. We substantiate this assertion further below by an in-depth analysis of the cascade dynamics.

The evolution of the average energy $\langle \varepsilon \rangle$, the dynamical quantum parameter $\langle \chi \rangle$, and the spatial distribution of the electrons is presented in Figs.~\ref{fig:chi_energy} and \ref{fig:density_single_pulse}. 
The S-type cascade sets in at $t\approx-0.4\tau_L$, when the initial electron approaches the strong field region [see Fig.~\ref{fig:density_single_pulse}a)] and $\langle \chi \rangle$ grows up to the values $\gtrsim 1$ (see Fig.~\ref{fig:chi_energy}). Later, as secondary particles are produced, $\langle \varepsilon \rangle$ and $\langle \chi \rangle$ rapidly drop until their values become insufficient to support the cascade, which thus collapses at $t\approx 0$. Eventually the particles get driven and turned around by the field. This can be seen from Fig.~\ref{fig:density_single_pulse}, where at $t\approx 0$ the electron spatial distribution broadens in the transverse direction. Furthermore, some secondary electrons reach the central area of the focus. In effect, such electrons become slow and can seed an A-type cascade.

After the S-type cascade collapses, if the laser field strength is sufficient, it restores $\langle \varepsilon \rangle$ and $\langle \chi \rangle$ of the slow electrons (see Fig.~\ref{fig:chi_energy}) and the pair production rate  starts growing again (see Fig.~\ref{fig:N}). This indicates the development of an A-type cascade. It saturates when, due to laser pulse diffraction, the EM field strength becomes insufficient to support it further.

The distribution of particles seeding the A-type cascade [see Fig.~\ref{fig:density_single_pulse}a) at  $t\approx 0$] differs from the idealized case considered in Sec.~\ref{sec:sp_estimates}. Namely, the number of such particles is large and they are distributed nonuniformly in the focal region. After the A-type cascade sets in, most of the particles are produced in the central region of the focus [see Fig.~\ref{fig:density_single_pulse}a) at  $t\gtrsim \tau_L$]. This means that the onset of A-type cascade is determined mainly by a small fraction of particles located near the optical axis (at $t\approx 0$). Therefore, Eqs.~ \eqref{eq:single_pulse_cond1} and \eqref{eq:single_pulse_cond2} give a reasonable order-of-magnitude estimate for the A-type cascade onset threshold even in a more realistic scenario considered here.

As the A-type cascade develops, the generated $e^-e^+\gamma$-plasma cushion co-propagates the laser pulse [see Fig.~\ref{fig:density_single_pulse}a)]. Therefore the suggested setup could be applied for generating short collimated dense bunches of high-energy electrons, positrons and/or photons. Density, energy spectrum and the divergence angle of these bunches could be controlled by tweaking the setup parameters.

Let us discuss the properties of a generated photon bunch. In Fig.~\ref{fig:gamma_pulse} we present the distribution of photons at the moment ($t=4.2\tau_L$) corresponding to the A-type cascade damping. The size of a photon bunch is approximately $4\lambda$ and $2\lambda$ in the transverse and longitudinal direction, respectively, where $\lambda$ is the laser pulse wavelength. 

Naturally, the form of the photon distribution is determined by the dynamics of the radiating electrons and positrons. They acquire transverse momentum from the field [see Fig.~\ref{fig:density_single_pulse}] as the laser pulse diverges, and also because they are pushed out of the region with a stronger electric field. This results in an angular spread of the photon distribution, see Fig.~\ref{fig:gamma_pulse}c), and explains why photons are partially concentrated in the electric field minima in Fig.~\ref{fig:gamma_pulse}a). The annular structures in Figs.~\ref{fig:gamma_pulse}b) and c) arise due to circular polarization of the field and resemble photon distributions produced by an interaction of a circularly polarised laser pulse of ultra-high intensity with a foil, see Ref. \cite{ji2014energy}. 

We present examples of photon energy spectra in Fig. \ref{fig:photon_spectra}. As we raise the peak laser field strength, they become wider and the mean photon energy  increases. This is peculiar to an A-type cascade (see Ref.~\cite{mironov2017observable}). Note that the maximal values of energy showing up in the spectra exceed the energy of the initial electron ($\varepsilon_0=2$ GeV). As the considered photons result from the A-type cascade, the width of the spectrum depends on the parameters of the field rather than on $\varepsilon_0$. The latter in general affects the multiplicity of the cascade.

\section{Summary and conclusions}
\label{sec:summary}
We have reconsidered a key distinctive feature of  selfsustained (A-type) QED cascades, the process of ongoing restoration of particles energy and dynamical quantum parameter after a hard photon emissions. Namely, by solving the classical equation of motion in terms of short-time expansion combined with ultrarelativistic approximation, we have identified the general short-term behavior of the energy and dynamical quantum parameter of an initially slow particle in an arbitrary electromagnetic field of electric type and established its validity conditions. 

Based on these results, we have generalized the previously proposed criteria for onset of selfsustained (A-type) QED cascades to an arbitrary electromagnetic field of electric type. The refined criteria are formulated in a local and Lorentz-invariant form (in terms of field invariants), reproduce the results discussed previously and quantify the required initial slowness of seed particles.

As an illustrative practical application, we have performed an in-depth analysis of an A-type cascade onset in a single focused laser pulse. By now, a systematic consideration of this fundamentally important and obviously most directly realizable setup remained lacking. For this case we have tested our refined general criteria against numerical simulations, first assuming a slow seed particle residing initially at the focal center. As the predicted dependence of the cascade onset threshold on focusing degree was found in a rather reasonable agreement with simulation results, this confirmed the qualitative rationality of our general approach and the results. 

Next we have considered a more realistic scenario, in which a cascade was seeded by a colliding head-on bunch of high-energy electrons. Initially the impact triggers an S-type cascade, but it was possible to tweak the setup parameters so that, after the particles loose their energy on emission and the initial S-type cascade fades out, they get driven by the laser pulse and an A-type one further sets in. The net multiplicity of so initiated A-type cascade is controlled by both the laser intensity $I$ and the energy $\varepsilon_0$ of electrons in the bunch. In our simulations, an A-type cascade developed with the setup parameters $\varepsilon_0=2$ GeV and $I\gtrsim 5\times10^{25}$ W/cm$^2$, when a $200$ PW laser pulse of optical frequency was focused to the diffraction limit ($\Delta=0.3$).

In this sort of setup, the	resulting A-type cascade creates a cushion of relativistic $e^-e^+$-plasma inside the pulse, and naturally acts as a converter of soft laser radiation into the high-energy photons. They are emitted in short bunches of duration comparable to the laser field period. Also, in contrast to the multi-beam setups, where radiation is emitted in various directions, here the emission is concentrated in a narrow forward cone, hence the secondary cascade can serve as a bright source of directed high-energy $\gamma$-quanta with tunable energy spectrum, luminosity and spatial divergence. 

The finding of this paper can be important for designing experiments at new laser facilities, aimed at studying the Strong Field QED phenomena, in particular, generating high-density relativistic $e^-e^+$ plasmas and high-energy photons.

\section*{Acknowledgements}
A.A.M. was supported by the Russian Foundation for Basic Research (Grant No. 19-32-60084) and the MEPhI Academic Excellence Project (Contract No. 02.a03.21.0005). 
E.G.G.  was supported by the project
ADONIS (Advanced research using high intensity laser produced photons and particles)
CZ.02.1.01/0.0/0.0/16\_019/0000789
from European Regional Development
Fund.
A.M.F. was supported by the MEPhI Academic Excellence Project (Contract No. 02.a03.21.0005), Russian Foundation for Basic Research (Grants No. 19-02-00643 and No. 20-52-12046), and the Tomsk State University Competitiveness Improvement Program.

\appendix
\needspace{3\baselineskip}
\section{Electron in a uniformly rotating electric field}
\label{sec:app_rot}
Let us illustrate the general arguments of Sec.~\ref{sec:chi(t)} by a simple explicit example. For a uniformly rotating electric field 
\begin{equation}
\label{eq:uniror_f}
\vec{E}(t)=\{E_0\cos(\omega t),E_0\sin(\omega t),0\},\quad\vec{H}=0,
\end{equation}
the solution for equation (\ref{eq:eq_motion}) with the initial condition $\vec{p}_0=\{0,p_y,0\}$ without any approximations takes the form
\begin{equation}\label{eq:ini_cond_tr}
\vec{p}(t)=\left\{ma_0\sin{\omega t},p_y+ma_0(1-\cos{\omega t}),0\right\},
\end{equation}
where $a_0=eE_0/m\omega$. Hence, by substituting Eq.~(\ref{eq:ini_cond_tr}) and $p_0\equiv \varepsilon(t)=\sqrt{\vec{p}^2(t)+m^2}$ into Eq.~(\ref{eq:chi}) and its further expansion in powers of $\omega t$ we obtain:
\begin{equation}
\begin{split}
\chi_e^2(t)=&\frac{e^2E_0^2\varepsilon_\perp^2}{m^6}-\frac{e^4E_0^4}{m^6}\frac{p_y}{ma_0}\left(1+\frac{p_y}{ma_0}\right)t^2\\
&+\frac{e^4E_0^4\omega^2}{12m^6}\left[3+7\frac{p_y}{ma_0}+4\left(\frac{p_y}{ma_0}\right)^2\right]t^4+\mathcal{O}(t^6),
\end{split}
\end{equation}
with an abbreviation $\varepsilon_\perp=\sqrt{p_y^2+m^2}$. Under the condition $p_y\ll ma_0$ [cf. Eq.~(\ref{eq:tr_mom}) and the discussion preceding its presentation] the terms containing the small ratio $p_y/ma_0$ can be neglected. Then the only remaining term of order $t^4$ is identical to the one given in Eq.~(\ref{eq:chi_t}) and can be obtained by exactly that prescription. Furthermore, as long as $t\gtrsim \sqrt{\varepsilon_\perp/eE_0\omega}$, it indeed exceeds both terms of orders $\mathcal{O}(1)$ and $\mathcal{O}(t^2)$, compare to the condition (\ref{eq:newcond}) and its derivation. In this explicit example one can also easily verify that the higher order terms indeed remain smaller as long as $t\ll 1/\omega$.

\color{black}
\section{Electron in a single focused laser pulse}
\label{sec:app_sp}
Let us specify the time dependence of the energy [Eq.~\eqref{eq:ultrarel_p}] and the parameter $\chi$ [Eq.~\eqref{eq:chi_t_short}] for an electron in a single laser beam. For definiteness assume that at  $t=0$ the electron is located at the focal center $\vec{r}=0$ with momentum $\vec{p}=0$. 

We employ a model of a focused laser beam suggested in \cite{narozhny2000scattering}. The EM field is formed by a superposition of plane waves, the resulting vector potential is given by 
\begin{equation}
\label{app:vec_pot}
\mathbf{A}(\mathbf{r},t)=\int \limits_{|\mathbf{k}'-\mathbf{k}|<\omega\Delta} d^3k' \mathbf{A}(\mathbf{k}')e^{i(\mathbf{k'r}-\omega t)},
\end{equation} 
where the wave vectors $\mathbf{k}'$ ($|\vec{k}'|=\omega$) fill a cone with an opening angle $\Delta$ around the carrier wave vector $\mathbf{k}$. The field strengths of a circularly $e$-polarized beam propagating along $z$-axis reads
\begin{widetext}
\begin{equation}\label{app:E_e}
{\bf{E}}^e=iE_0  e^{-i\varphi}\left\{ F_1({\bf e}_x\pm i{\bf
e}_y)-F_2 e^{\pm 2 i\phi}({\bf e}_x\mp i{\bf e}_y)\right\},
\end{equation}
\begin{equation}\label{app:H_e}{\bf{H}}^e=\pm E_0 e^{-i \varphi}
\left\{\left(1-i\Delta^2\frac{\partial}{\partial\mathcal{Z}}\right)
\left[F_1({\bf e}_x\pm i{\bf e}_y)+F_2 e^{\pm 2 i\phi}({\bf
e}_x\mp i{\bf e}_y)\right]+2i\Delta e^{\pm i\phi}\frac{\partial
F_1}{\partial \mathcal{R}}{\bf e}_z \right\}.
\end{equation}
\end{widetext}
Here we use the notations
\begin{equation}
\begin{array}{c}
\varphi=\omega(t-z),~~~\mathcal{R}=\rho/R,~~~\mathcal{Z}=z/L,  \\ \\
\rho=\sqrt{x^2+y^2},~~~\cos\phi=x/\rho,~~~\sin\phi=y/\rho, \\ \\
\Delta\equiv 1/\omega R=\lambda/2\pi R,~~~L\equiv R/\Delta\,.
\end{array}
\end{equation}
As implied in Eq.~\eqref{app:vec_pot}, the model was initially formulated with a step-like aperture having support on a cone $|\mathbf{k}'-\mathbf{k}|<\omega\Delta$. However, assuming small opening angles $\Delta\ll 1$, it is more convenient to replace it with a Gaussian one $\mathbf{A}(\mathbf{k}')\propto \exp(-\mathbf{k}'^2/\omega^2\Delta^2)$ and extend the limits of integration to the whole $\mathbf{k}'$-space. Then the functions $F_{1,2}$ in Eqs.~\eqref{app:E_e} and \eqref{app:H_e} can be expressed explicitly as
\begin{equation}\label{app:F_funcs}
\begin{array}{c}
\displaystyle
\begin{split}
F_1=&(1+2i\mathcal{Z})^{-2}\left(1-\frac{\mathcal{R}^2}{1+2i\mathcal{Z}}\right) \\
&\times\exp\left(-\frac{\mathcal{R}^2}{1+2i\mathcal{Z}}\right),
\end{split}\\ 
\displaystyle
F_2=-\mathcal{R}^2(1+2i\mathcal{Z})^{-3}\exp\left(-\frac{\mathcal{R}^2}{1+2i\mathcal{Z}}\right)\,.
\end{array}
\end{equation}

The EM tensor $F^\mu_{\;\;\;\nu}(x)$ is worked out straightforwardly. In what follows, we only need the values of $F^\mu_{\;\;\;\nu}$ and its derivative at $x^\mu=0$. The explicit expression for $F^{(0)\mu}_{\quad\;\;\;\nu}=F^\mu_{\;\;\;\nu}(0)$ is given by
\begin{equation}
\label{app:F0}
F^{(0)\mu}_{\quad\;\;\;\nu}=E_0
\begin{pmatrix}
\mqty{ 0 & 0 & -1 & 0 \\
 0 & 0 & 0 & 0 \\
 -1 & 0 & 0 & 1-4\Delta^2 \\
 0 & 0 & -1+4\Delta^2}
\end{pmatrix}.
\end{equation}

According to Sec.~\ref{sec:chi(t)}, in order to obtain the first order corrections $p^{(1)}$ and $x^{(1)}$ to the solution of the equations of motion, we first need to solve the eigenvalue problem $F^{(0)}f_i=\lambda_i f_i$, $i=1,\,2,\,3,\,4$.  The eigenvalues of the matrix \eqref{app:F0} are given by
\begin{equation}
\label{app:EV}
\lambda_i = 2\sqrt{2}E_0\Delta \sqrt{1 - 2\Delta^2}\times\lbrace 1,\, -1,\, 0,\, 0 \rbrace
\end{equation}
and the corresponding eigenvectors by
\begin{equation}
\label{app:f_i}
\begin{array}{l}
f_1^\mu = \left(1,\,0,\, -2 \sqrt{2}\Delta \sqrt{1 - 2 \Delta^2},\, 1 - 4 \Delta^2 \right),\\
f_2^\mu = \left(1,\, 0,\, 2 \sqrt{2}\Delta \sqrt{1 - 2 \Delta^2},\, 1 - 4 \Delta^2 \right),\\
f_3^\mu = \left(0,\, 1,\, 0,\, 0 \right),\\
f_3^\mu = \left(1-4\Delta^2,\, 0,\, 0,\, 1 \right),
\end{array}
\end{equation}
respectively. Note that the eigenvectors $f_{1,2}^\mu$ are normalized so that $f_{1,2}^0=1$ (see Sec. \ref{sec:chi(t)}). To construct a general solution $p^{(1)}$ we also need to identify the constants $C_i$ from the initial conditions. By plugging \eqref{app:f_i} into Eq.~\eqref{eq:1_order_sol} at $\tau=0$, we obtain:
\begin{equation}
\label{app:C_i}
C_i=\frac{m}{16\Delta^2(1-2\Delta^2)} \lbrace 1,\,1,\,0,\,2(1-4\Delta^2) \rbrace.
\end{equation}

Now let us stick to our approximation. As $\epsilon=\lambda_1$, we immediately obtain the energy of the electron $\varepsilon(t)\approx e\epsilon t$, see Eq.~\eqref{eq:main1}. To find the dependence $\chi(t)$, according to Eq.~\eqref{eq:chi_t_short}, we need to work out explicitly the expression
\begin{equation}
\label{app:omega_eff}
\epsilon^2\omega_\mathrm{eff}= \epsilon^2\sqrt{f_{1\mu}F^\mu_{\;\;\;\nu,\sigma}f_1^\sigma\left( J^{-1}\right)^{\nu}_{\;\;\;\lambda}F^\lambda_{\;\;\;\varkappa,\rho}f_1^\rho f_1^\varkappa},
\end{equation}
where $J=4\epsilon^2-F(0)^2$. The main building block is the 
combination $F^\mu_{\;\;\;\nu,\sigma}(0)f_1^\sigma$, which reads

\begin{widetext}
\begin{equation}
\label{app:Ff}
F^\mu_{\;\;\;\nu,\sigma}(0)f_1^\sigma = 8 E_0\omega\Delta^2
\begin{pmatrix}
\mqty{ 0 & 1-2\Delta^2 & 0 & 0 \\
 1-2\Delta^2 & 0 & -2\sqrt{2}\Delta\sqrt{1-2\Delta^2} & -1+7\Delta^2-12\Delta^4 \\
 0 & 2\sqrt{2}\Delta\sqrt{1-2\Delta^2} & 0 & 0 \\
 0 & 1-7\Delta^2+12\Delta^4 & 0 & 0}
\end{pmatrix}.
\end{equation}
The full expression for the inverse of $J$ is rather cumbersome, but for $\Delta\ll 1$ with the same accuracy as above can be simplified to
\begin{equation}
\label{app:J}
(J^{-1})^{\mu}_{\;\;\;\nu}=\frac{1}{768 E_0^2\Delta^4}
\begin{pmatrix}
\mqty{ 1+28\Delta^2+60\Delta^4 & 0 & 0 & -1+4\Delta^4 \\
0 & 24\Delta^2+48\Delta^4 & 0 & 0 \\
 0 & 0 & 32\Delta^2+64\Delta^4 & 0 \\
 1-4\Delta^4 & 0 & 0 & -1+28\Delta^2+52\Delta^4}
\end{pmatrix}+\mathcal{O}(\Delta^2).
\end{equation}
\end{widetext}
After substitution of $f_1$, Eq.~(\ref{app:Ff}) and Eq.~(\ref{app:J}) into Eq.~(\ref{app:omega_eff}), up to the leading order in $\Delta$ we obtain:
\begin{equation}
\epsilon^2\omega_\mathrm{eff}=136\sqrt{2}E_0^2\omega\Delta^5+\mathcal{O}(\Delta^5),
\end{equation}
thus finally arriving at Eq.~(\ref{eq:main2}).


%

\end{document}